\documentclass[namedreferences]{solarphysics}

\usepackage[hyperref,optionalrh,showbiblabels]{spr-sola-addons} 
\usepackage{color}           
\usepackage{breakurl}        
\usepackage{hyperref}
\usepackage{hhline}
\usepackage[dvipsnames]{xcolor}
\usepackage{booktabs}
\usepackage{gensymb}
\usepackage{multirow}
\usepackage{listings}
\usepackage{rotating}
\usepackage{lipsum}
\usepackage{epstopdf}

\usepackage{tikz}
\usetikzlibrary{spy}

\chardef\us=`\_

\begin{document}

\begin{article}
\begin{opening}

\title{Review of image quality measures for solar imaging\\ {\it Solar Physics}}

\author[addressref={aff1},corref,email={apopowicz@polsl.pl}]{\inits{A.}\fnm{Adam}~\lnm{Popowicz}}
\author[addressref=aff1]{\inits{K.}\fnm{Krystian}~\lnm{Radlak}}
\author[addressref=aff2]{\inits{K.}\fnm{Krzysztof}~\lnm{Bernacki}}
\author[addressref=aff3]{\inits{V.G.}\fnm{Valeri}~\lnm{Orlov}}


\address[id=aff1]{Institute of Automatic Control, Silesian University of Technology, Akademicka 16, Gliwice, Poland}
\address[id=aff2]{Institute of Electronics, Silesian University of Technology, Akadamicka 16, Gliwice, Poland}
\address[id=aff3]{Instituto de Astronom{\'i}a,Universidad Nacional Autonoma de M{\'e}xico, Apdo. Postal 70-264, Cd. Universitaria, 04510 M{\'e}xico D.F., M\'{e}xico}

\runningauthor{Author-a et al.}
\runningtitle{Example paper}

\begin{abstract}
The observations of solar photosphere from the ground encounter significant problems due to the presence of Earth's turbulent atmosphere. Prior to applying image reconstruction techniques, the frames obtained in most favorable atmospheric conditions (so-called lucky frames) have to be carefully selected. However, the estimation of the quality of images containing complex photospheric structures is not a trivial task and the standard routines applied in nighttime Lucky Imaging observations are not applicable. In this paper we evaluate 36 methods dedicated for the assessment of image quality which were presented in the rich literature over last 40 years. We compare their effectiveness on simulated solar observations of both active regions and granulation patches, using reference data obtained by the Solar Optical Telescope on the Hindoe satellite. To create the images affected by a known degree of atmospheric degradation, we employ the Random Wave Vector method which faithfully models all the seeing characteristics. The results provide useful information about the methods performance depending on the average seeing conditions expressed by the ratio of the telescope's aperture to the Fried parameter, $D/r_0$. The comparison identifies three methods for consideration by observers: Helmli and Scherer's Mean, Median Filter Gradient Similarity, and Discrete Cosine Transform Energy Ratio. While the first one requires less computational effort and can be used effectively virtually in any atmospherics conditions, the second one shows its superiority at good seeing ($D/r_0<4$). The last one should be considered mainly for the post-processing of strongly blurred images.
\end{abstract}
\end{opening}

\section{Introduction}
Ground-based telescopes suffer from the degradation of image quality due to the turbulent nature of Earth's atmosphere. This phenomenon, frequently termed as ''seeing'', prevents large-aperture telescopes from achieving their theoretical angular resolution. Even the best observing sites do not allow for observations in the visible with the resolution higher then the diffraction limit of a 20 cm telescope. This means, that long exposures of both very small and extremely large telescope show the same angular resolution.



The Fried parameter, $r_0$, is a quantity which describes the average atmospheric conditions. The Fried parameter is the distance across which the expected change of the wavefront phase is exactly 1/2$\pi$. It may be also understood as the size of the telescope's aperture over which the theoretical diffraction limit may be easily achieved. Since at the best observational sites $r_0$ rarely reaches 20~cm in the visible ($\mathrm{\lambda=0.5~\mu m}$) \citep{SolarSites}, the long exposures obtained from any telescope do not expose the resolution higher than the one reached by a 20 cm telescope. Importantly, the $D/r_0$ ratio is frequently utilized to determine how far the resolution of images obtained by a telescope of $D$ diameter is from its theoretical diffraction limit. 

So far, numerous techniques, both hardware- and software-based, have been developed to enhance the resolution of astronomical observations. Probably the most prominent hardware-based approach is adaptive optics \citep[AO,][]{AO1}, in which the compensation of wavefront distortions is performed directly by a deformable mirror. The AO for solar imaging differs from the one used in nighttime observations \citep{SolarAO1,SolarAO2}. There is no point-like object for wavefront sensing which is performed usually with Shack-Hartmann sensor. Instead, the cross-correlation between images observed in individual sub-apertures has to be calculated to estimate the shape of wavefront \citep{SolarAO3}. Significantly poorer daytime atmospheric conditions put much higher demands on the updating frequency of deformable mirror. Fortunately, there is also more light available for sensors which enables such observations.

A popular example of software-based approach to improve the quality of images is the method called Lucky Imaging \citep[LI,][]{Scharmer1989,LI1}. Due to the availability of very fast and low-noise cameras, this method has recently become very popular high-resolution acquisition technique in the visible. However, it is dedicated to smaller telescopes ($<$ 2 m), since it relies on the fact, that there is only a small chance to obtain a diffraction-limited image in a series of very short exposures (i.e., shorter than the coherence time of the atmosphere -- usually up to several milliseconds). This chance, estimated by \cite{LI2}, is relatively high for smaller apertures (e.g., $D/r_0<2$), while it quickly becomes negligible for greater mirrors.

There are many combinations of software and hardware techniques of high-resolution imaging. As an example, \cite{AOLI1} or \cite{AOLI2} present the fusion of Lucky Imaging and adaptive optics (AOLI). In \citet{holo} the deconvolution of a series of short exposures is shown as another possible way to enhance the quality of astronomical images.  A wide range of speckle-interferometry methods is also available \citep{saha,SI1}. They involve such approaches as the aperture masking \citep{AM1,AM2} or speckle bispectral reconstructions \citep{speckleMasking,bispectrum}. Some of them have been successfully used for a long time in solar imaging \citep{SolRecOld}.


Either with or without AO, the observer acquires a series of images and then applies several post-processing steps \citep[e.g.,][]{solar_postproc}. One of them is always the selection of best exposures, i.e., the ones having the highest image quality. Such lucky imaging has been very popular for a long time in ground-based solar observatories due to its apparent simplicity and very low hardware requirements (only a fast camera is necessary) \citep{LIsolarOLD}. However, the most advanced imagers utilizing lucky exposures are far from being simple \citep{DKIST}. The rejection of less useful frames is possible due to the high intensity of observed object which allows for acquiring thousands of exposures per second at relatively low resolution or tens of frames if large-format cameras are utilized. The selection is also essential for reaching the high quality of final outcomes regardless of the complexity of succeeding image reconstruction (e.g., simple stacking, deconvolution, or bi-spectral analysis).

The assessment of the temporal quality of registered solar images becomes challenging due to the complex character of observed scene. It is impossible to utilize a basic quality metric frequently used in lucky imaging -- the intensity of the brightest pixel in a speckle pattern -- as it requires a well isolated point-like object (star). Instead, a widely employed method for quality assessment of solar images is the root-mean-square contrast (rms-contrast) which assumes the uniform isotropic properties of granulation \citep{rms1,solar_postproc,rms2}. Unfortunately, the method has several drawbacks like a significant dependency of its effectiveness on the wavelength \citep{rms_bad1} and the sensitivity to the structural contents of the image \citep{DengZHang2015}. This implies that also the tip-tilt effects, which move observed patch slightly changing its contents, will introduce additional noise to quality estimation as the image features, like sunspots or pores, will move in and out of the analyzed region.

Motivated by a recent work by \cite{DengZHang2015} introducing an objective image quality metric to solar observations, we decided to explore which of the numerous quality metrics (QMs) available in the rich literature can be employed for selection of solar frames. This is carried out by investigating the correlation between the outcomes of QMs and the known strength of simulated turbulence. Our review includes 36 QMs with many varying implementations. Implementations refers to different gradient operators, kernel sizes, thresholding techniques, etc., without implying conceptual changes. We utilize reference images from the Solar Optical Telescope (SOT) on board the Hindoe satellite \citet{Tsuneta2008} and use an advanced method for modeling atmospheric turbulences, the Random Wave Vectors \citep[RWV,][]{RWV1} method. The scintillation noise is also included to faithfully reflect all the seeing characteristics. Moreover, we check the computational efficiency of the QMs to indicate which methods are more suitable for application in high-speed real-time image processing.

\section{Experiment}
\subsection{Turbulence simulation}
\begin{figure}
\centering
\includegraphics[width=0.5\linewidth]{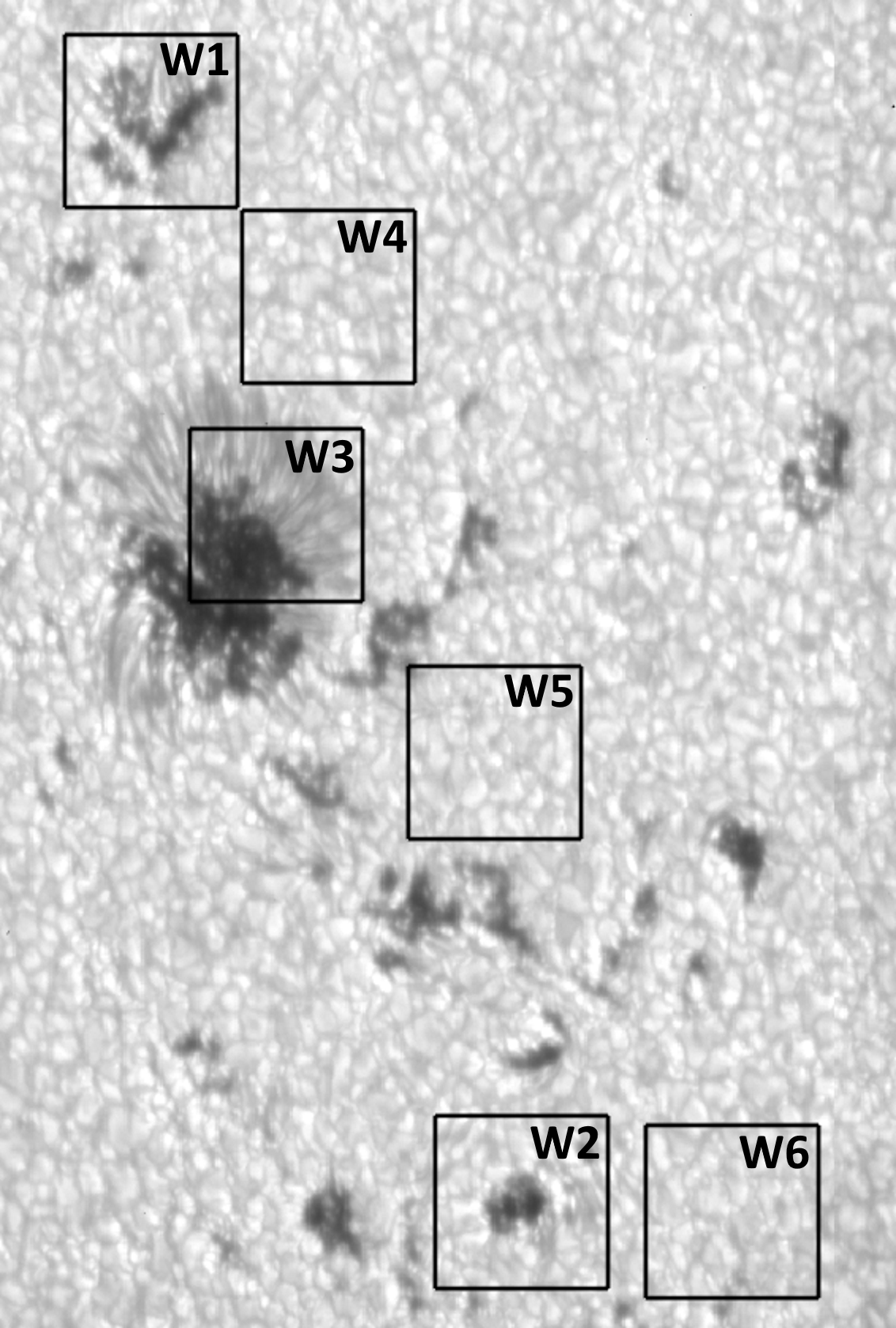}
\caption{A part of solar photosphere imaged by the Hindoe satellite. The six patches, ($W1-W6$, $100\times100$ pixels) extracted for the experiment, are annotated. Three of them ($W1,W2,W3$) contain the parts of active regions, while the remaining ones ($W4,W5,W6$) include only granulation. The image was obtained in the green continuum on 29 Nov. 2015 at 21:19:44 UT.}
\label{allSun}
\end{figure}

Since the observations from the space are not disturbed by the atmosphere, as the reference data for our experiments we used the images obtained by the SOT. From a wide range of registered images in Hindoe database\footnote{HINDOE database query form is available at \url{http://darts.isas.jaxa.jp/solar/hinode/query.php} }, we selected a one which contained several regions of enhanced magnetic activity (sunspots). The image was originally obtained at green continuum wavelength, on 29 Nov. 2015 at 21:19:44UT. In the Fig. \ref{allSun} we show the selected reference and indicate the positions of six extracted $100\times100$-pixel patches, (roughly $5'' \times 5''$ each, image scale of 0.054"~pixel$^{-1}$). Patches W1, W2, and W3 include sunspots, while W4, W5, and W6 contain granulation. 

\begin{figure*}
\centering
\includegraphics[width=\linewidth]{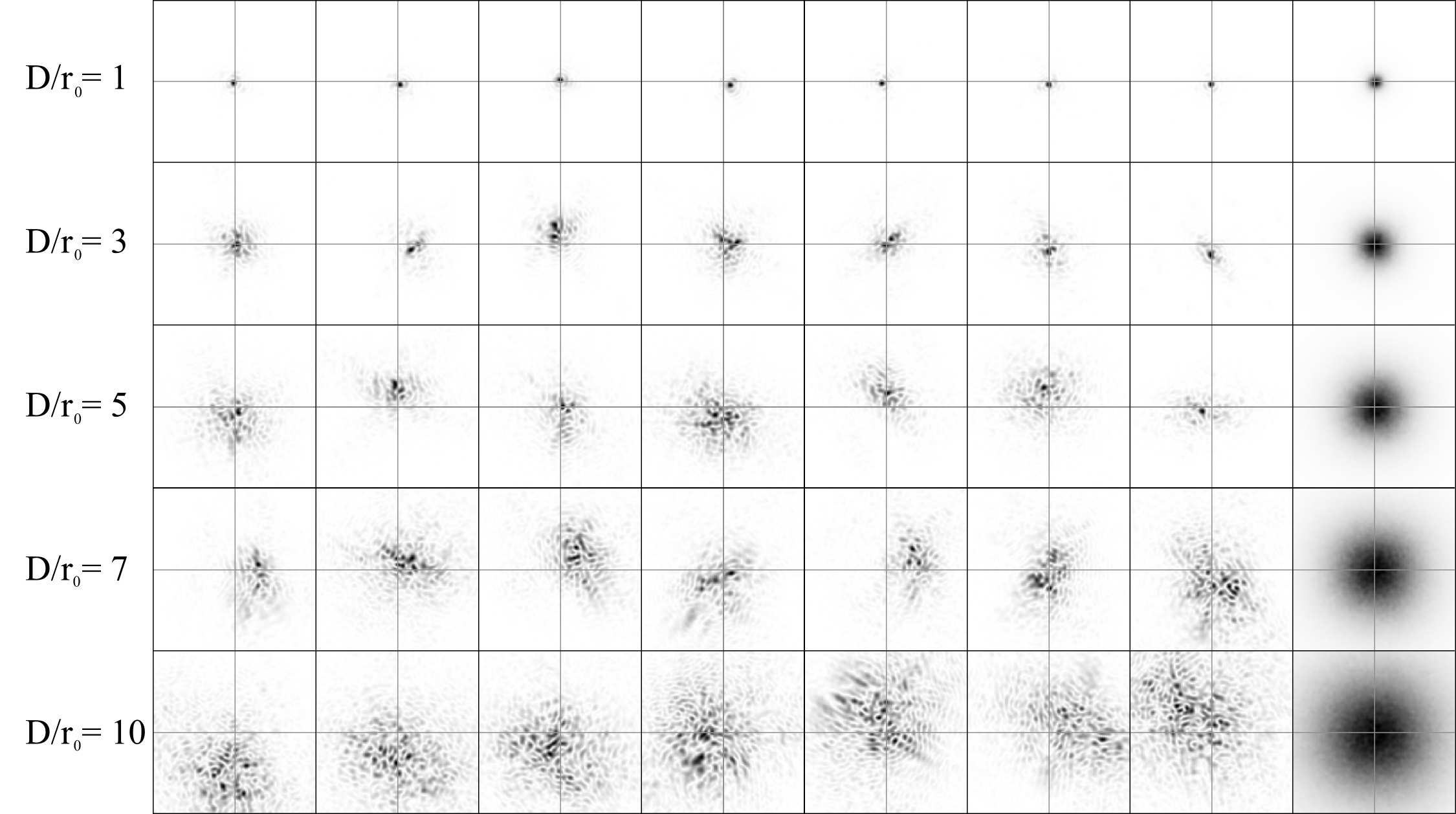}
\caption{Exemplary blurring kernels (negatives) employed in the experiment. The rows correspond to various $D/r_0$ relative atmospheric conditions while the last column exposes the simulated long-exposure seeing disk (i.e., an ensemble average over all the kernels for a given $D/r_0$). Each kernel is presented in a box with a side length of $40\lambda/D$. The auxiliary gray lines indicate the box center highlighting the simulated tip-tilt effect.}
\label{kernels1}
\end{figure*}

To investigate the response of various QMs for a given strength of turbulence, we modeled the transfer function of the atmosphere utilizing RWV method. The method allows for reliable modeling of amplitude and phase fluctuations of the incoming optical wavefront. Since a patch of solar photosphere, which is used for quality assessment, can be arbitrarily small we assumed that it is within the isoplanatic angle. Thus, anisoplanatic effects, more complicated for simulation, are neglected in this study, which will otherwise significantly complicate the simulations. Using RVW we generated 1000 blurring kernels (speckles patterns) for ten distinctive scenarios of seeing conditions: $D/r_0$ = 1, 2, ..., 10. We assumed that such a range is representative for observations at best observing sites. Each generated kernel consisted of 1024$\times$1024 pixels and the resolution was two times higher than required by the Nyquist limit, i.e, a single pixel corresponded to an angular size of $\lambda / 4D$ ($\lambda$ -- wavelength, $D$ -- telescope diameter). We opted for such oversampling to be able to combine simulated speckle kernels with reference Hindoe images. Each blurring kernel was normalized so that the summed intensity over all pixels is unity. Exemplary kernels with the corresponding long-exposure seeing disks are presented in Fig. \ref{kernels1}. Evidently, the poorer the seeing conditions (higher $D/r_0$), the more complex is the kernel shape. The simulated tip-tilt effect is also visible as a displacement of the kernel's centroid.

Atmospheric scintillation results in varying attenuation of flux collected by a telescope. This type of noise was recently investigated by \cite{osborn}. The authors present the following formula for estimating the relative variance of the total intensity of the observed object:
\begin{equation}
\sigma_{s} =\sqrt{10^{-5}~C^2_Y~D^{-4/3}~t_e^{-1}~ (\textrm{cos} \gamma)^{-3}~\textrm{exp}(-2h_{\mathrm{obs}}/H)},
\end{equation}
where $D$ is telescope diameter, $\gamma$ is the zenith distance , $h_{\mathrm{obs}}$ is the altitude of the observatory, $H$ is the scale height of the atmospheric turbulence, generally assumed to be $\sim$ 8000 m, $C_Y$ is the scaling coefficient which can be determined from turbulence profilers (SCIDARs) and was estimated between 1.30$-$1.67 for best observing sites (see Tab. 1 in the original work \citet{osborn}). 

To include such fluctuations in our simulated images, we multiplied each kernel by a random variable with an expected value of unity and a standard deviation equal to $\sigma_{s}$. We assumed (1) the telescope size $D=0.5$ m according to the size of Hindoe/SOT instrument, (2) the observations at zenith distance of $\gamma=60^{\circ}$ and (3) a low scintillation index, $C_y=1.5$, which is expected for (4) high-altitude observatory, $h_{\mathrm{obs}}= 3000$ m. For such conditions the relative scintillation noise is $\sigma_s = 0.032$. 

\begin{figure}
\centering
\includegraphics[width=0.5\linewidth]{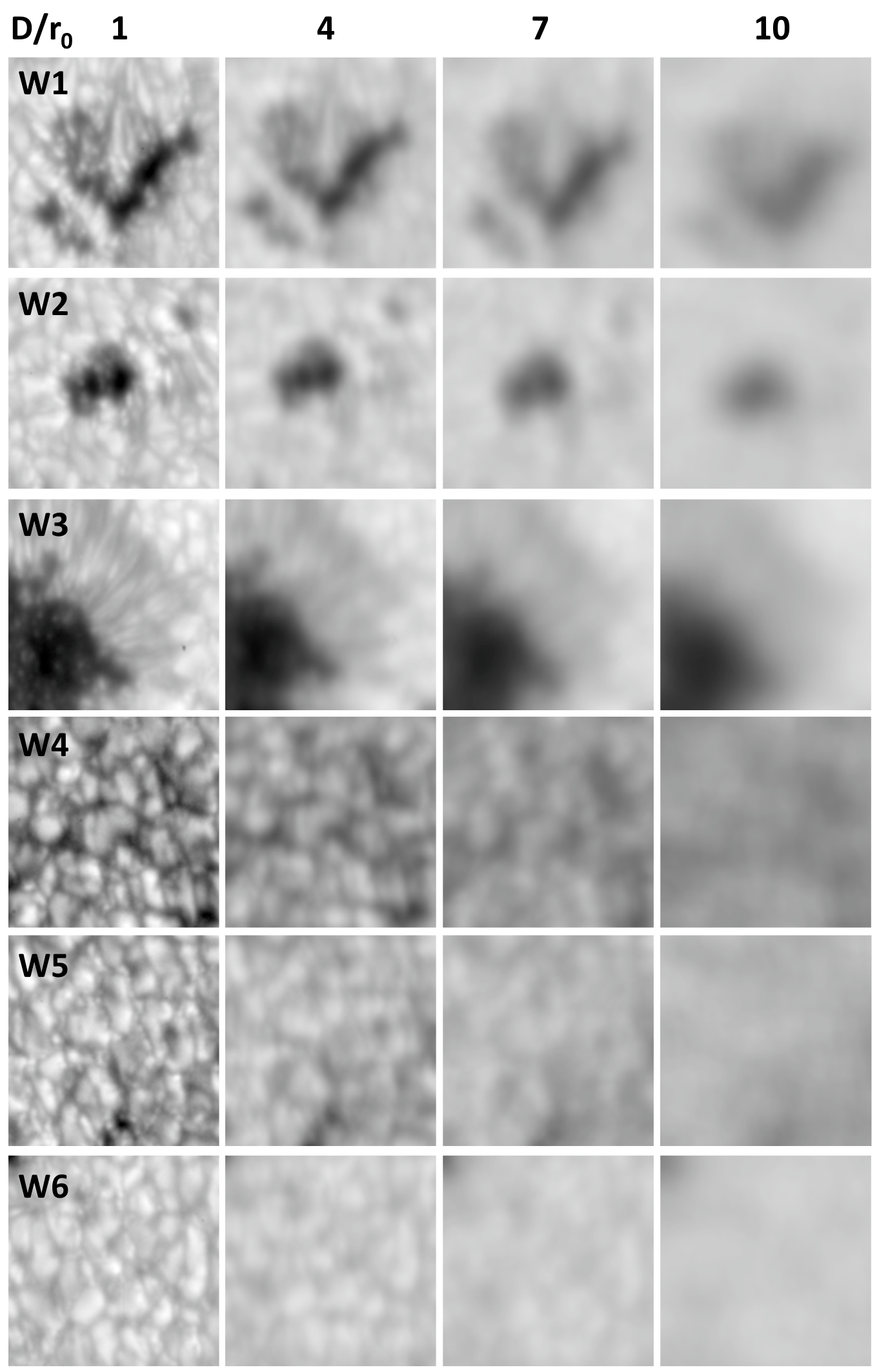}
\caption{Exemplary simulated images for all six patches (rows) and four selected degrees of atmospherics turbulence, $D/r_0=1,~4,~7,\textrm{~and~}10$ (columns).}
\label{patches1}
\end{figure}

To properly convolve a blurring kernel with a reference patch, the scale of a kernel has to be the same as the image scale. For the assumptions given above, i.e., $D = 0.5$ m telescope observing at $\lambda= 550$ nm, we obtain an image scale of 0.055'' pixel$^{-1}$. This means that the sampling of blurring kernels ($D/4
\lambda$) and the assumed telescope size allow for convolving kernels with the Hindoe images without any prescalling. Several examples of solar images degraded by simulated blurring kernels are presented in Fig. \ref{patches1}.

The time-dependent quality, when observing stellar objects, can be determined from the relative intensity of the brightest pixel in the normalized kernel. This is a widely accepted approach to select the sharpest frames in the nighttime lucky imaging \citep{LI1}. However, in our case the object is not point-like and shows complex structures. Thus, we decided to use the quality measure based on the amount of energy preserved from the original frequency spectrum. Since the original image is convolved with a blurring kernel and the original amplitudes in frequency spectrum are multiplied by the amplitudes of a kernel, the value of proposed quality measure can be calculated by summing squared amplitudes in 2-D Fourier transform of a kernel. According to the Parseval's theorem, this also equals to the sum of squared intensities of a kernel directly in the image plane.

\begin{figure}
\centering
\includegraphics[width=\linewidth]{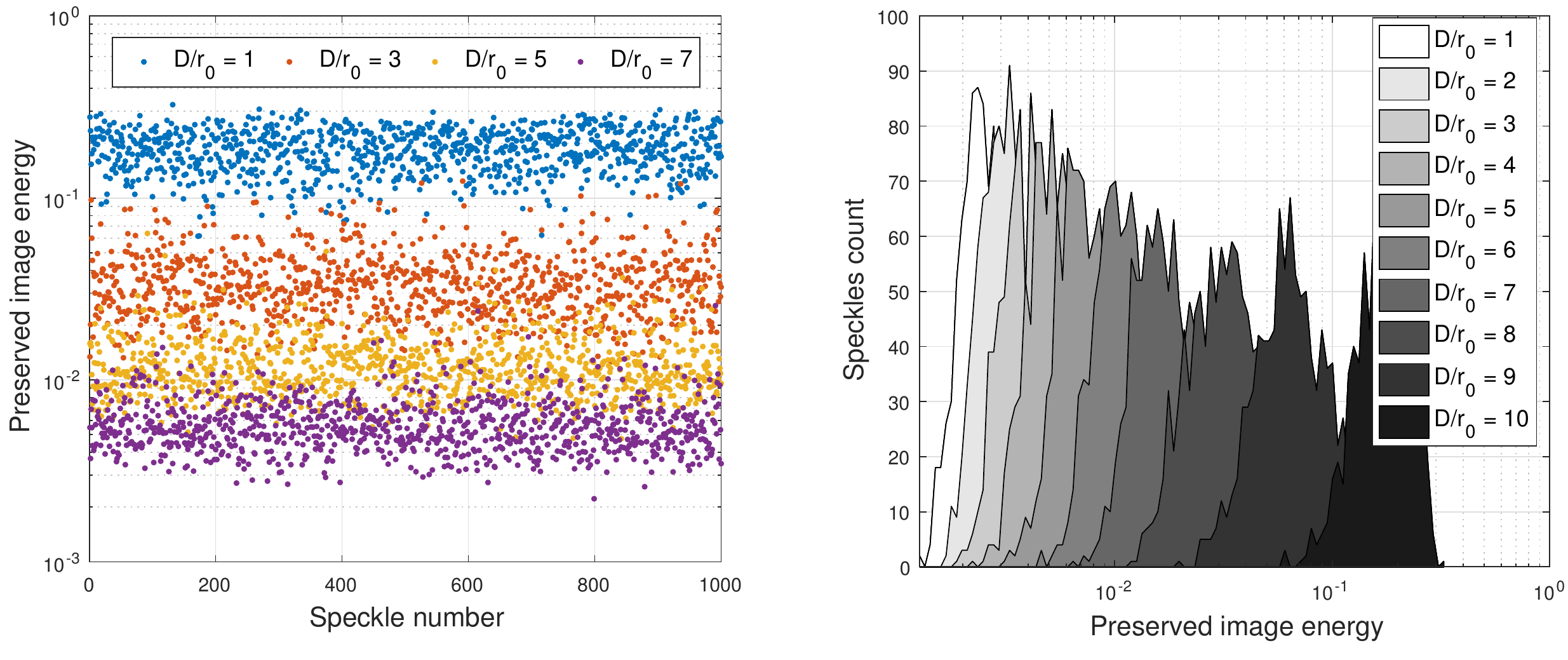}
\caption{Dependencies of temporal turbulence strength expressed by the maximal intensity in normalized blurring kernel, on the average seeing conditions $D/r_0$. \emph{Left}: amount of preserved image energy for three 1000-kernel sets $D/r_0=1,4$, and 10. \emph{Rigth}: histograms of the preserved image energy over all the simulated kernels. (Color figures are available in online version.)}
\label{qKernel}
\end{figure}

Since turbulence is a random process, there is also a possibility that the temporal seeing conditions will become much better or much worse than the ones indicated by the average $D/r_0$ \citep{LI2}. This is in fact what we also observed in our data. For all ten sets of simulated kernels ($D/r_0=1,2,...,10$), the spread of temporal quality is exposed in the histograms in Fig. \ref{qKernel}. In the upper part of Fig. \ref{qKernel} we plotted the amount of preserved energy of the original image over all 1000 kernels for three values of $D/r_0$. Clearly, the quality for the average $D/r_0=4$ can sometimes outperform the conditions registered at $D/r_0=1$. Therefore, $D/r_0$ expresses only the average blurring strength while it can not be taken as a reliable estimator of the quality of individual frames.

\subsection{Methods}
A set of 35 state-of-the-art QMs was provided in the review of \citet{PertuzPuig2013}. The authors consider the most popular methods dedicated for the assessment of focus in complex scenes by means of contrast measurement. The methods can be categorized into six families of operators, based on: gradients (GRA), Laplacians (LAP), wavelets (WAV), intensity statistics (STA), discrete-cosine transform (DCT), and the miscellaneous methods (MIS), i.e., the ones that can not be categorized into any other group. An overview of the methods with their abbreviations and references is given in Tab.~\ref{tab_par_ref}. 

Within the set of techniques one can find the most popular method used in solar image processing, i.e., rms-contrast measure, which was called Normalized Gray-level Variance and abbreviated as STA5 (we followed this nomenclature). The most recent QM proposed by \cite{DengZHang2015}, Median Filter Gradient Similarity (GRA8), was included in our comparison as well. 
\begin{table*}
\centering
\caption{Overview of quality measures.}
\begin{tabular}{lll}
 \hline 
 Abbr.      & Operator name       &  Citation    \\
 \hline
 \multicolumn{3}{c}{ \bf{ Miscellaneous operators (MIS*)}} \\ \hline
 
 MIS1&    Absolute Central Moment     &  \citet{Shirvaikar2004}\\
  MIS2&   Brenner's Focus measure     &  \citet{BrennerDew1976}\\ 
   MIS3&    Image Contrast     &  \citet{Nanda2001}\\
    MIS4&    Image Curvature measure    &  \citet{HelmliScherer2001}\\
     MIS5&    Helmli and Scherer's Mean     &  \citet{HelmliScherer2001}\\
      MIS6&    Local Binary Patterns measure    &  \citet{LorenzoCastrillon2008}\\
       MIS7&    Steerable Filters-based measure     &  \citet{MinhasMohammed2009}\\
        MIS8&   Spatial Frequency measure     &  \citet{MalikChoi2008}\\
         MIS9&    Vollath's Autocorrelation   &  \citet{Vollath1987}\\
    \hline
\multicolumn{3}{c}{ \bf{ Gradient-based operators (GRA*)}} \\ \hline
  GRA1    & Gaussian Derivative     & \citet{GeusebroekCornelissen2000} \\
   GRA2    & Gradient Energy     & \citet{SubbaraoChoi1993} \\
    GRA3    & Thresholded Gradient      & \citet{SantosOrtiz1997,KuangChern2001} \\

     GRA4    & Squared Gradient     & \citet{GeusebroekCornelissen2000} \\
      GRA5  &3D Gradient     & \citet{AhmadChoi2007} \\
       GRA6  & Tenengrad     & \citet{Tenenbaum1971,SchlagSanderson1983}   \\ 
          GRA7    & Tenengrad Variance    & \citet{PechPacheco2000}  \\ 
      GRA8      &  Median Filter Gradient Similarity      & \citet{DengZHang2015}     \\
          \hline
  
  \multicolumn{3}{c}{ \bf{ Laplacian-based operators (LAP*)}} \\ \hline
      
   LAP1 &   Energy of Laplacian     & \citet{SubbaraoChoi1993}     \\
   LAP2 &      Modified Laplacian    & \citet{NayarNakagawa1994}     \\
   LAP3 &       Diagonal Laplacian   & \citet{Thelen2009}     \\
   LAP4 &     Variance of Laplacian     & \citet{PechPacheco2000}      \\ \hline    
      
       \multicolumn{3}{c}{ \bf{ Wavelet-based operators (WAV*)}} \\ \hline
      WAV1 &     Sum of Wavelet Coefficients   & \citet{YangNelson2003} \\
      WAV2 &      Variance of Wavelet Coefficients  & \citet{YangNelson2003} \\
    WAV3 &     Ratio of Wavelet Coefficients   & \citet{XieRong2006} \\ 
     WAV4 &      Ratio of Curvelet Coefficients  & \citet{MinhasMohammed2009}          \\
    \hline  
      
       \multicolumn{3}{c}{ \bf{ Statistics-based operators (STA*)} } \\ \hline
 STA1 &      Chebyshev Moments-based  & \citet{YapRaveendran2004}     \\
 STA2 &     Eigenvalues-based   & \citet{WeeParamesran2007}     \\
 STA3 &      Gray-level Variance  & \citet{GroenYoung1985}     \\
 STA4 &      Gray-level Local Variance  & \citet{PechPacheco2000}     \\
 STA5 &      Normalized Gray-level Variance  & \citet{GroenYoung1985}     \\
 STA6 &      Modified Gray-level Variance  & \citet{PertuzPuig2013}     \\
 STA7 &      Histogram Entropy  & \citet{FirestoneCook1991}     \\
 STA8 &      Histogram Range  & \citet{FirestoneCook1991}     \\
      \hline  
    
     \multicolumn{3}{c}{  \bf{ DCT-based operators (DCT*)}} \\ \hline
      
   DCT1 &      DCT Energy Ratio  & \citet{ShenChen2006}          \\
   DCT2 &     DCT Reduced Energy Ratio   & \citet{LeeYoo2009}          \\
   DCT3 &      Modified DCT  & \citet{LeeKumar2008}          \\
 \hline

\end{tabular}
\label{tab_par_ref}
\end{table*}

To improve the effectiveness of QMs for solar observations we had to adjust the parameters of many techniques. This was carried out experimentally by tuning the parameter considering the properties of observed solar scene and analyzing the algorithm's details. Moreover, for some of the methods we proposed major, but still simple, modifications which allowed for enhancing their effectiveness. This resulted in creation of various implementations of the most of methods (labeled as version $A$, $B$, etc.). The set of investigated parameter values and/or the details of applied modifications are given in Tab. \ref{method_params}. The distance parameters, like radius or size of local filtering window, are expressed in pixels (in our case 1 pixel = 0.055''). The thresholds are given in relative values, which means that before applying thresholding the intensities in a patch were normalized such that the intensities cover the range from zero to unity. Several methods have two adjustable parameters, therefore we evaluated various combinations of their values. Including all the modifications, we end up in a total number of 105 implementations of 36 techniques which are included in our comparison.

\begin{table*}
\centering
\caption{The parameters and modifications of the quality measures compared in the experiment. Some methods do not have any parameters, thus they were left blank.}
\label{method_params}
\begin{tabular}{lll}
 \hline 
 Method      & Variants   &  Description  \\
 \hline
       
\multicolumn{3}{c}{ \bf{ Miscellaneous operators (MIS*)}} \\ \hline
 
 MIS1&    A$-$D     &  number of bins in a histogram of image intensities ($L$)\\
& & $L=\{ 50,100,256, 1000\}$\\
MIS2&     A    &   only horizontal gradient \\
&     B    &  only vertical gradient\\
&   C     &  maximum of vertical and horizontal gradients \\
& D$-$G  & sum of vertical and horizontal gradients with thresholding ($T$)\\
& &  $T=\{0,0.01,0.1,0.5\}$\\

MIS3&    --     & --  \\
MIS4&    --     & --  \\
MIS5&   A$-$C     & window size ($w$)\\
& & $w=\{ 3,7,15\}$\\
MIS6&   A$-$D       & radius of the circle in LBP operator ($r$)  \\
&          & number of equally spaced pixels on the circle ($n$)   \\
& & $(r,n)=\{(1,8),(1,16),(2,8),(2,16) \}$\\
MIS7&  A$-$H     & size of the Gaussian mask  ($N$)\\
&  & standard deviation of the Gaussian ($\sigma$) \\
&  &  $(N,\sigma)=\{(7,2),(7,5),(7,10),(7,15),(15,2),(15,5),$\\
& & ~~~~~~~~~~~~$(15,10),(15,15) \}$ \\
MIS8$-$9&    --     & -- \\
\hline
\multicolumn{3}{c}{ \bf{ Gradient-based operators (GRA*)}} \\ \hline
GRA1    & A$-$F &    size of Gaussian mask ($N$)   \\
     &  & standard deviation of the Gaussian ($\sigma$)\\
     &  & $(N,\sigma)=\{ (7,2),(7,5),(7,10),(15,2),(15,5),(15,10)\}$\\

GRA2    &   --     & --   \\
GRA3    &   A    & maximum of  the horizontal and vertical gradients   \\
    &    B$-$F    &  horizontal gradient bigger than a threshold ($T$)   \\
    & &  $T=\{ 0.0,0.01,0.05,0.1, 0.5\}$ \\
 &    G$-$K   & horizontal and vertical gradients bigger than a threshold ($T$)\\
 & &  $T=\{ 0.0,0.01,0.05,0.1, 0.5\}$ \\  
    GRA4$-$7    &    --    & -- \\
 GRA8     &   A     & horizontal gradient and constant window size ($w$=3) \\
GRA8     &   B    & vertical gradient and constant window size ($w$=3)       \\
  GRA8     &   C$-$F    & horizontal and vertical gradient and window size $w$ \\
  & &     $w=\{ 3,5,7,11\} $  \\
       
         \hline
 
 \multicolumn{3}{c}{ \bf{ Laplacian-based operators (LAP*)}} \\ \hline
     
  LAP1$-$4 &   --     &  -- \\
  \hline    
      
       \multicolumn{3}{c}{ \bf{ Wavelet-based operators (WAV*)}} \\ \hline
     WAV1 &  A &1-level DWT with Daubechies-6 filters   \\
       & B  &2-level DWT and Daubechies-10 filters    \\
      WAV2 &  A & 1-level DWT with Daubechies-6 filters    \\
      & B & 2-level DWT and Daubechies-10 filters     \\
    WAV3$-$4&     --   & -- \\ 
   \hline

\end{tabular}
\end{table*}

\begin{table*}
\centering
\caption{Continuation of Tab. \ref{method_params}}
\label{method_params2}
\begin{tabular}{lll}
 \hline 
 Method      & Variants   &  Description  \\
 \hline
      
      \multicolumn{3}{c}{ \bf{ Statistics-based operators (STA*)} } \\ \hline
 STA1 &    A$-$E    &  order of Tchebichef polynomials ($p$) \\
 & &  $p=\{2,3,4,5,6\}$    \\
STA2 &    A$-$E     & number of diagonal elements from the matrix of eigenvalues  ($k$) \\
& &  $k=\{ 1,3,5,7,9\}$   \\
 STA3 &   --     &  -- \\
 STA4 &   A$-$E   & size of the window in which the local variance is computed ($w$)\\
 & &    $w=\{ 3,5,7,11,15\}$  \\
 STA5 &  --      & --   \\
STA6 &   A$-$E      & size of the window in which the local mean is computed ($w$) \\
& &   $w=\{ 3,5,7,11,15\}$   \\
STA7 &     A$-$D    & number of bins in histogram of image intensities  ($L$)\\
& &      $L=\{ 50,100,256,1000\}$\\
 STA8 &  A$-$D     &  percent of pixels removed from the histogram ($\alpha$)\\
 & &  $\alpha=\{ 0,0.01,0.02,0.05\}$   \\
   \hline  
    
   \multicolumn{3}{c}{  \bf{ DCT-based operators (DCT*)}} \\ \hline
     
   DCT1 & A$-$C  & size of sub-block of the image      \\
   & &   $w=\{ 4,8,16\}$      \\
   DCT2 &  A$-$C  &  size of sub-block of the image    \\
   & &    $w=\{ 4,8,16\}$       \\
   DCT3 &   --     &  --    \\
   \hline
 \end{tabular}
\end{table*}

\subsection{Data analysis}

To assess the performance of all methods we investigated the correlation between the results of QMs and the actual expected quality $Q$. As stated before, the quality was estimated by the total preserved energy with respect to the reference, undisturbed image. We observed that the relation between these two quantities is almost always nonlinear. Fortunately, it is sufficient that the dependency is monotonic as it allows us to distinguish between better and worse atmospheric conditions. Therefore, to estimate the effectiveness of the methods, instead of Pearson's correlation coefficient, we used the Spearman rank-order correlation $C_s$. Such a coefficient is insensitive to any nonlinearities in observed dependencies. 



\begin{figure*}
\centering
\includegraphics[width=\linewidth]{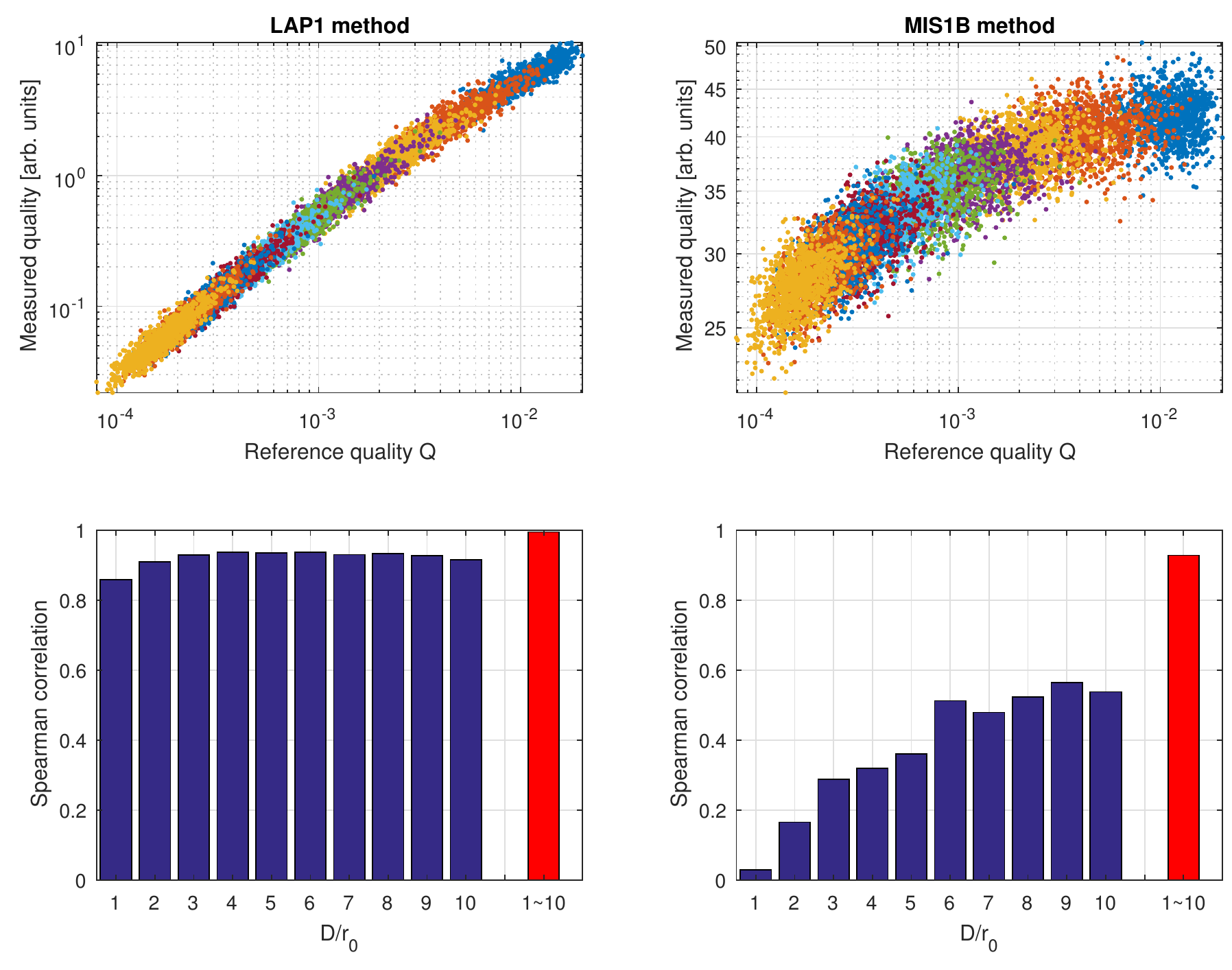}
\caption{Examples of correlation between measured and expected image quality for two distinct methods: \emph{LAP1} -- left side and \emph{MIS1B} -- right side. The analysis was performed for image patch W1. The calculated correlation coefficients in each $D/r_0$ set and over whole data set ($D/r_0=1-10$) are presented in the lower bar plots. Colors in point correspond to various $D/r_0$ conditions. (Color figures are available in online version.)}
\label{corrExamples}
\end{figure*}

As an example in the upper panel of Fig. \ref{corrExamples} we show significantly different dependencies between the expected quality Q and the outcomes of two QMs (\emph{LAP1} and \emph{MISB}, on the left and right side, respectively) calculated for image patch $W1$. Each set of $D/r_0$ was presented with a different marker/color. We decided to calculate the Spearman correlation coefficient in two ways: (1) within a set of observations for each $D/r_0$ and (2) across all the simulated conditions $D/r_0=1-10$. We present the corresponding results of $C_s$ in each $D/r_0$ subset in the bar plots of Fig. \ref{corrExamples}. The correlation depends on $D/r_0$ so that it becomes apparent which range of atmospheric conditions is most suitable for a given method. For example, the LAP1 method allows for effective estimation of image quality in $D/r_0=1$ while MIS1B is completely useless in this range. The proposed approach to calculate the correlation coefficient permitted estimating the performance of each method for a range of typical atmospheric conditions. 

In the example presented in Fig. \ref{corrExamples} the superiority of LAP1 over MIS1B is also visible for $C_s$ calculated over whole set of $D/r_0$. However, the difference is significantly smaller than in particular groups as both methods achieve $C_s > 0.9$ (0.98 and 0.90 for LAP1 and MIS1B, respectively). This is due to the wide range of considered $Q$ values, which makes the correlation between the quantities more evident and leads to the elevated value of the $C_s$ coefficient. Although the differences are not so significant, the correlation for such a wide range of $Q$ values still allows for comparison of the overall effectiveness of QMs, however, within much narrower range of $C_s$ values.

\subsection{Results and discussion}
\begin{figure}
\centering
\includegraphics[width=\linewidth]{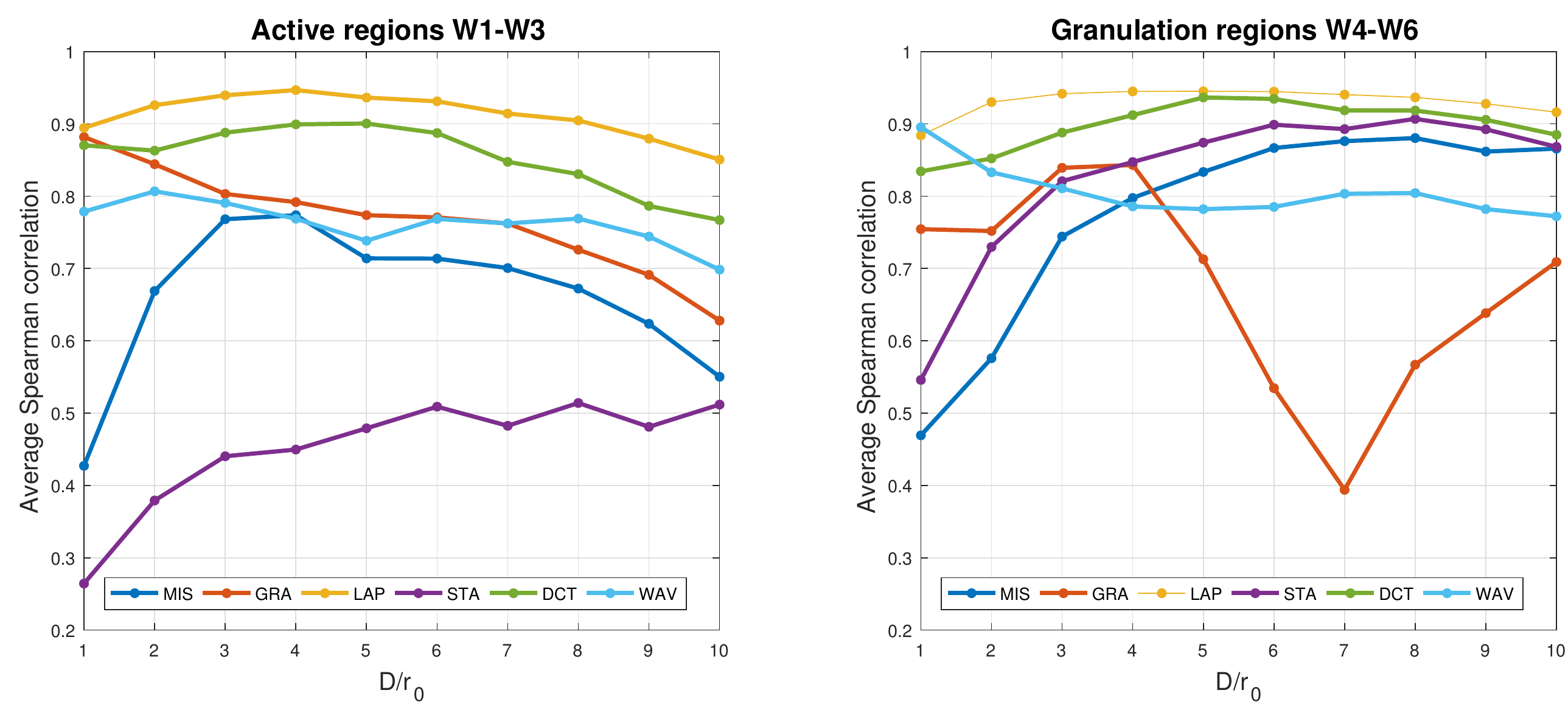}
\caption{Average correlation coefficients obtained for QM families. The left panel corresponds to the results for active regions, while the right one shows outcomes for granulation patches. (Color figures are available in online version.)}
\label{families}
\end{figure}

The results of the comparison are presented in Tab. \ref{tab_wyniki} and Fig. \ref{families}. While in Tab.~\ref{tab_wyniki} we present only the best four methods for each $D/r_0$, the average performance obtained for whole QM families is summarized in Fig. \ref{families}. We divided the results into two categories, according to the contents of observed image: W1, W2, and W3 -- active regions, and W4, W5, and W6 -- granulation. 

\begin{table*}
\setlength\tabcolsep{1.5pt}
\caption{The rankings of best quality measures for various $D/r_0$ seeing conditions and six analyzed patches: W1, W2, W3 -- active regions and W4, W5, W6 -- granulation areas. The lower bar plots show the rankings for wide-range correlation accross all atmospheric conditions $D/r_0=1-10$.}
\begin{tabular}{llccccc}
\hline

& & $D/r_0=1$ & $D/r_0=2$ & $D/r_0=3$ & $D/r_0=4$ & $D/r_0=5$ \\ \hline 

 \raisebox{2.8\normalbaselineskip}[1.85cm][0pt]{\rotatebox[origin=c]{90}{ W1--W3 }} &
  \raisebox{2.8\normalbaselineskip}[1.85cm][0pt]{\rotatebox[origin=c]{90}{ active regions }} &\includegraphics[width=0.175\textwidth]{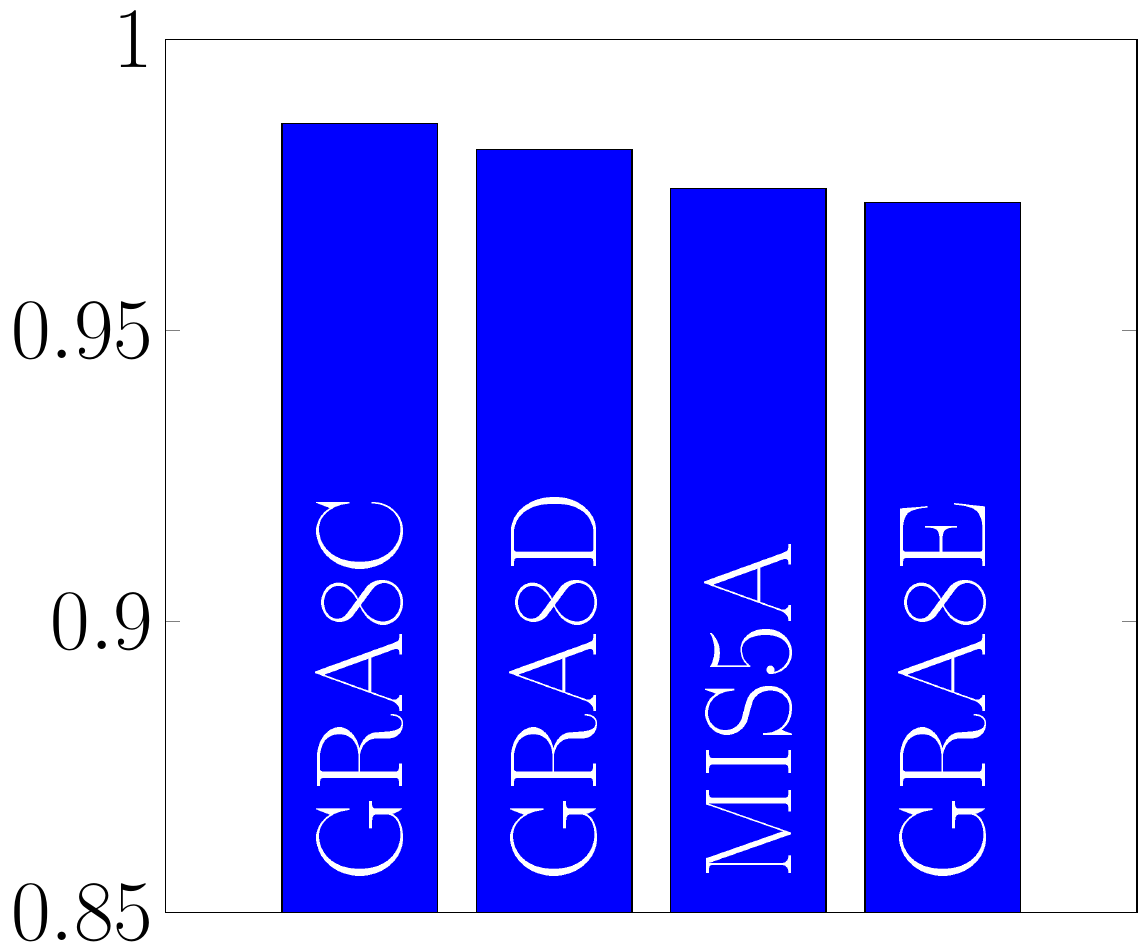} & \includegraphics[width=0.175\textwidth]{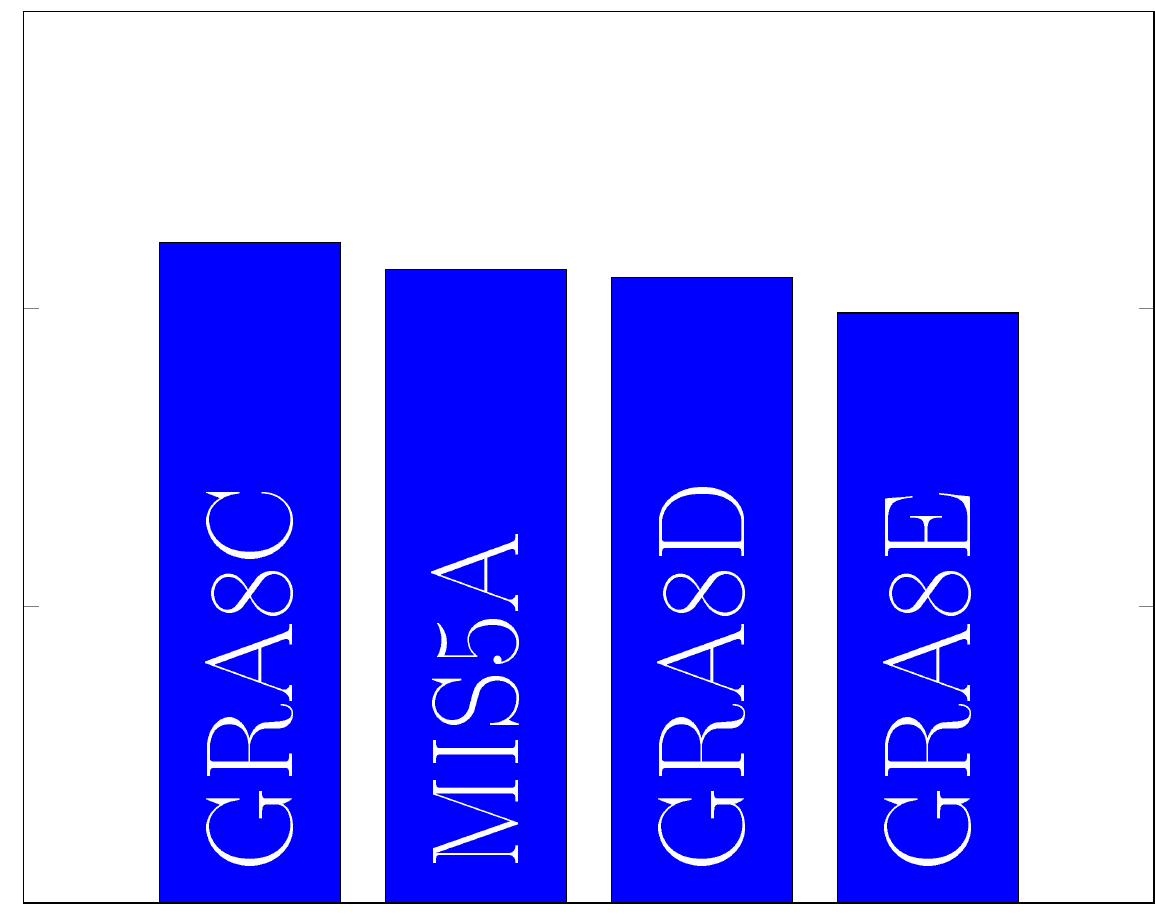} & 
\includegraphics[width=0.175\textwidth]{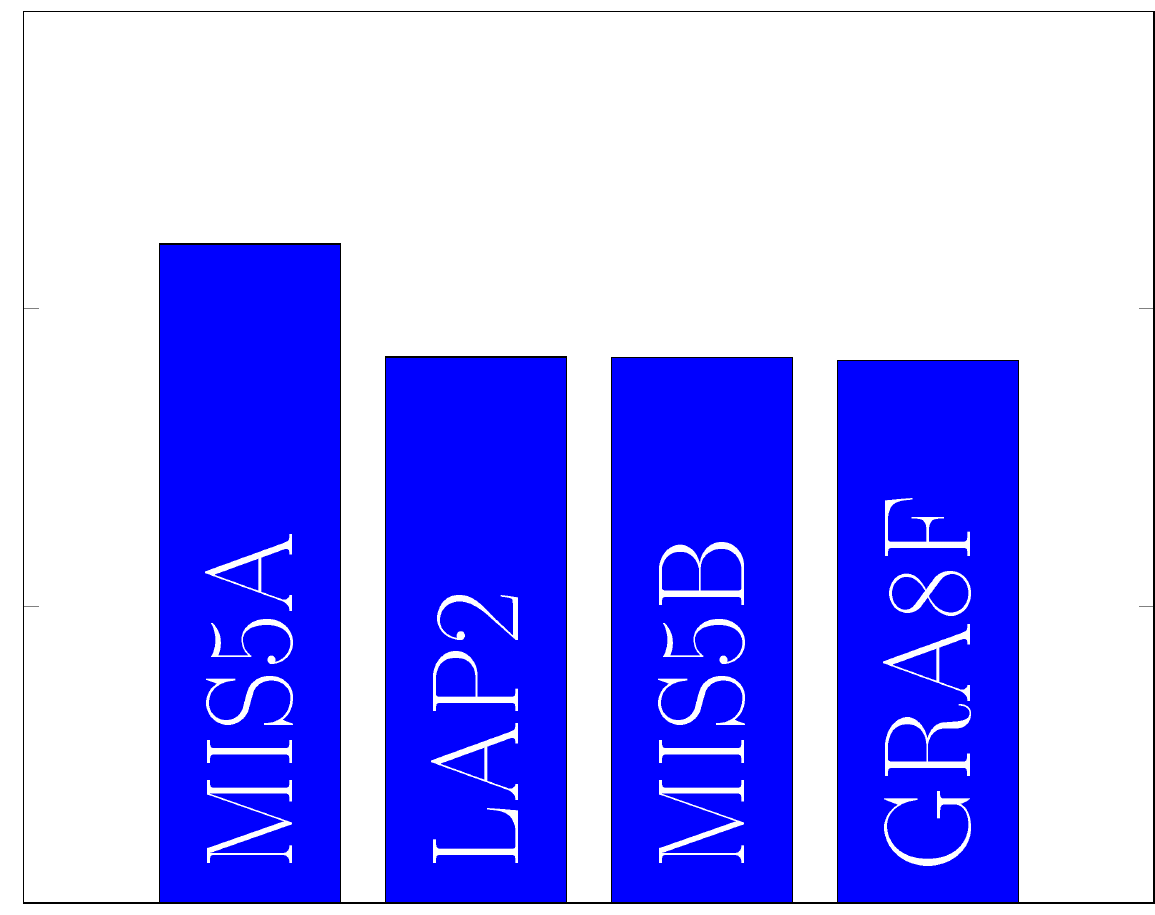} & 
\includegraphics[width=0.175\textwidth]{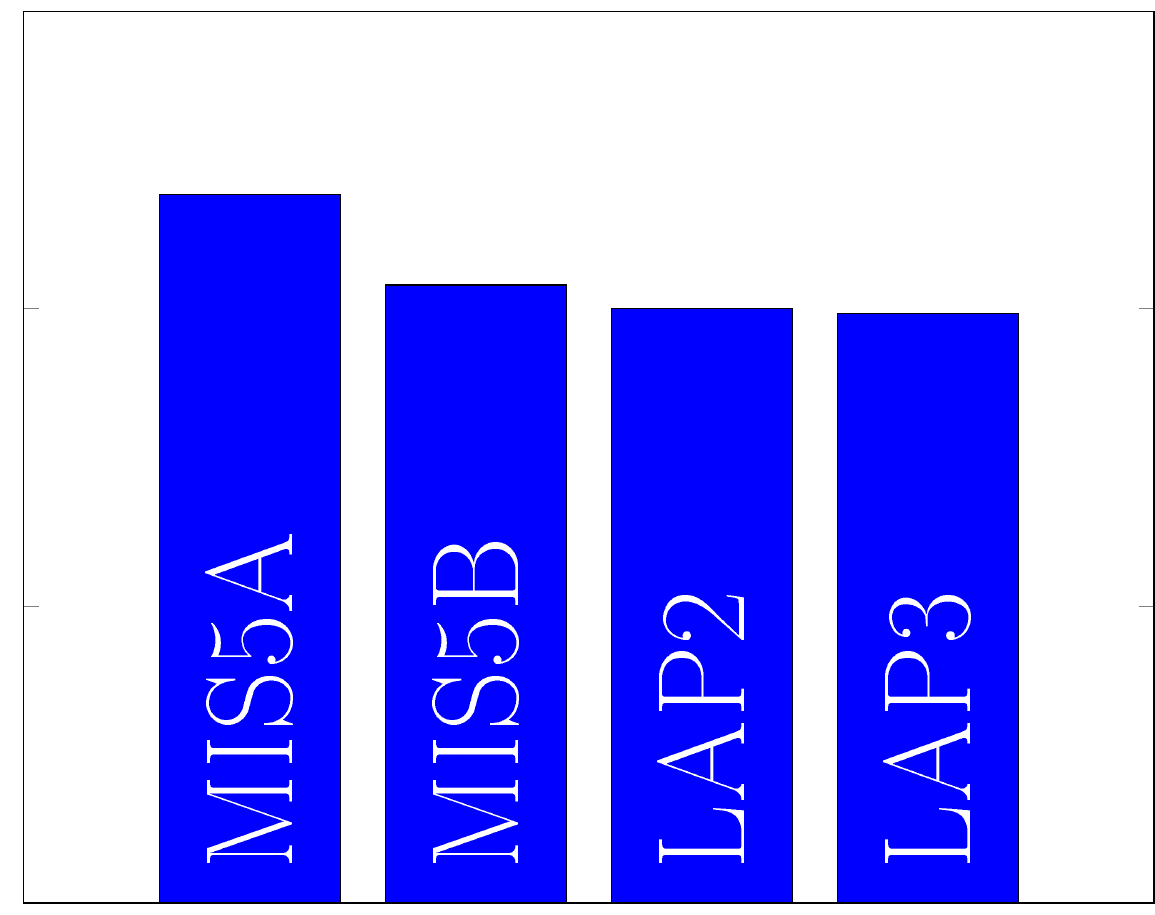} & 
\includegraphics[width=0.175\textwidth]{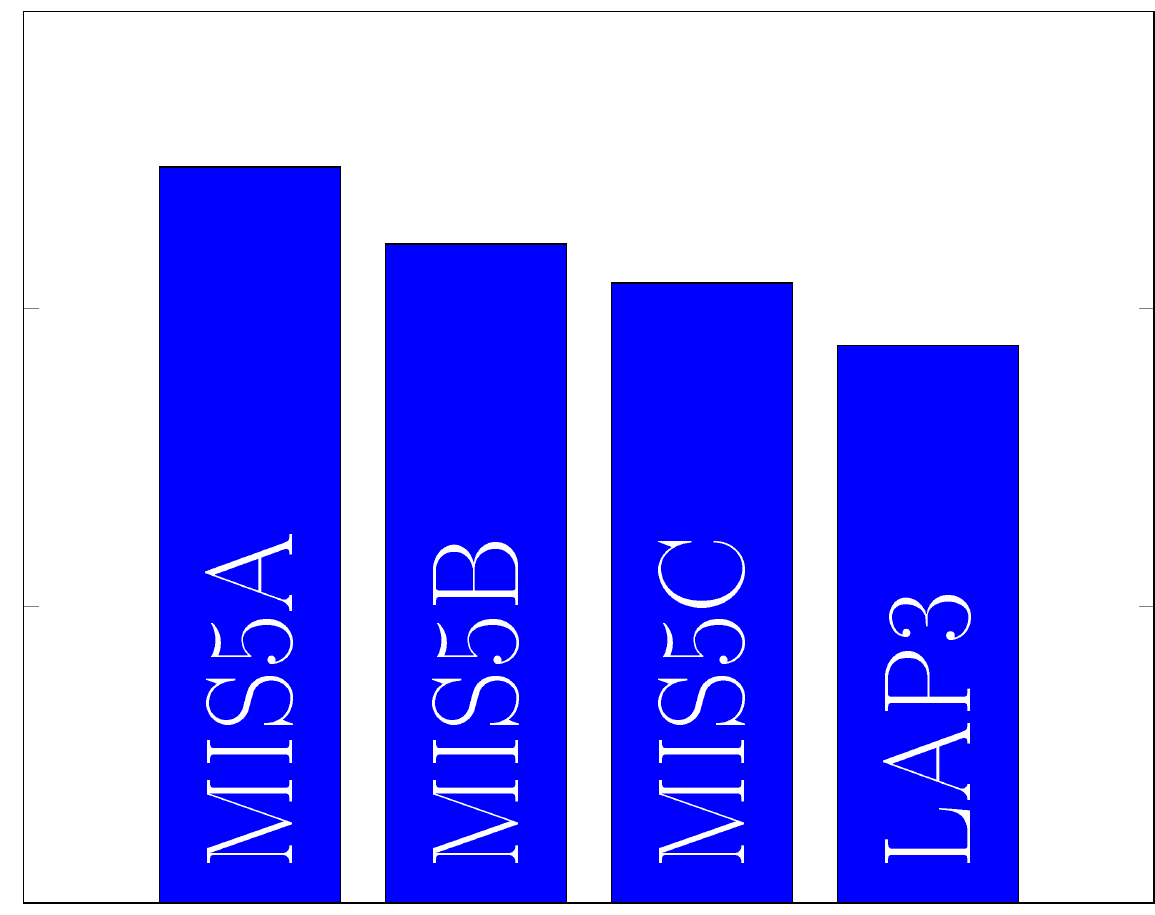} \\
 \raisebox{2.8\normalbaselineskip}[1.85cm][0pt]{\rotatebox[origin=c]{90}{ W4--W6}}  &
 \raisebox{2.8\normalbaselineskip}[1.85cm][0pt]{\rotatebox[origin=c]{90}{ granulation}}  &
\includegraphics[width=0.175\textwidth]{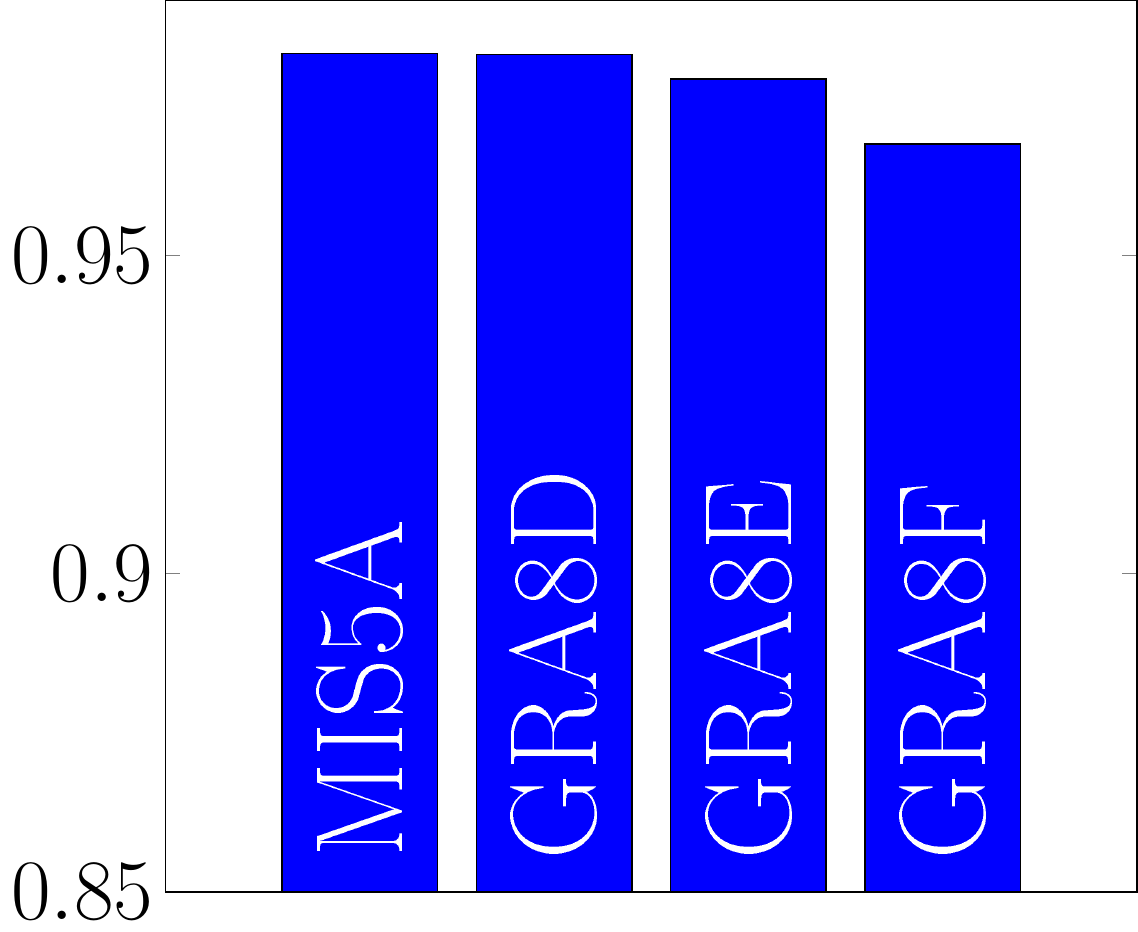} &  \includegraphics[width=0.175\textwidth]{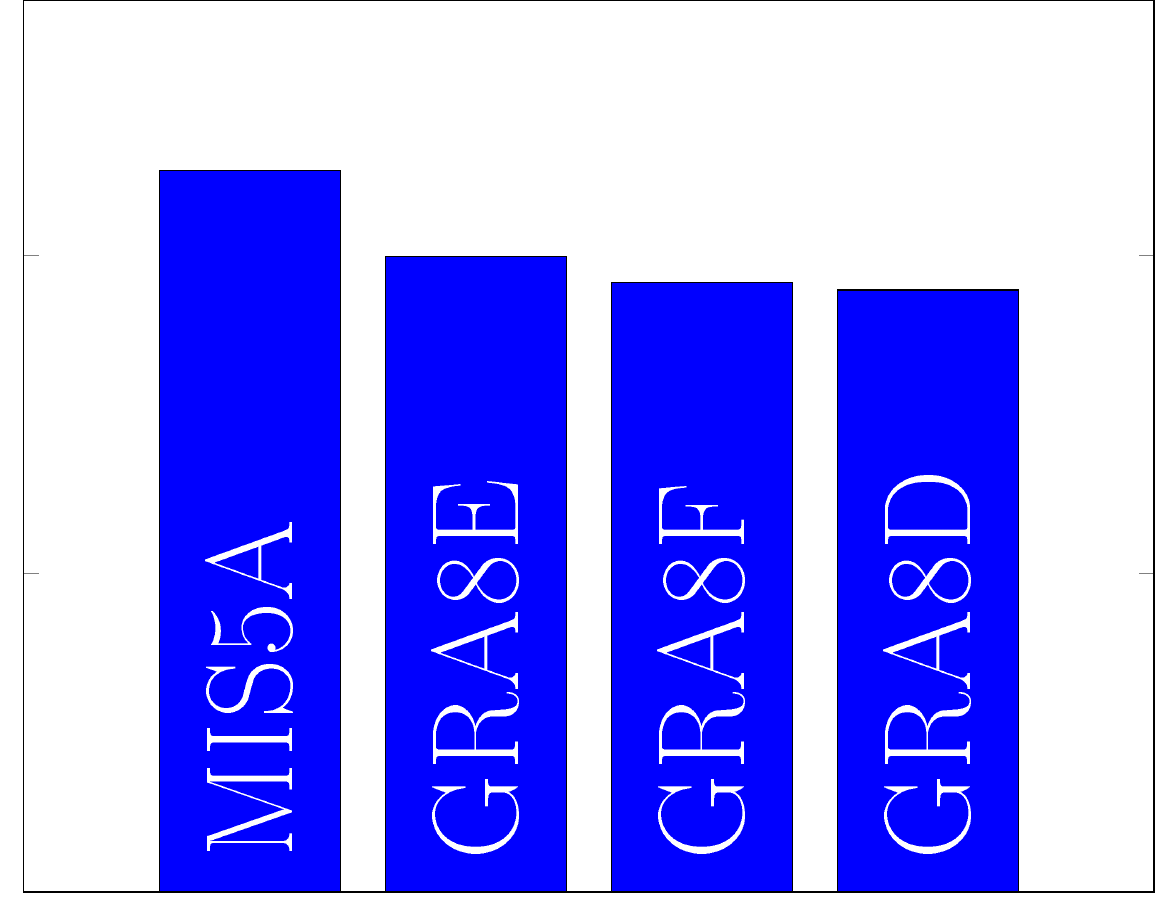} & 
\includegraphics[width=0.175\textwidth]{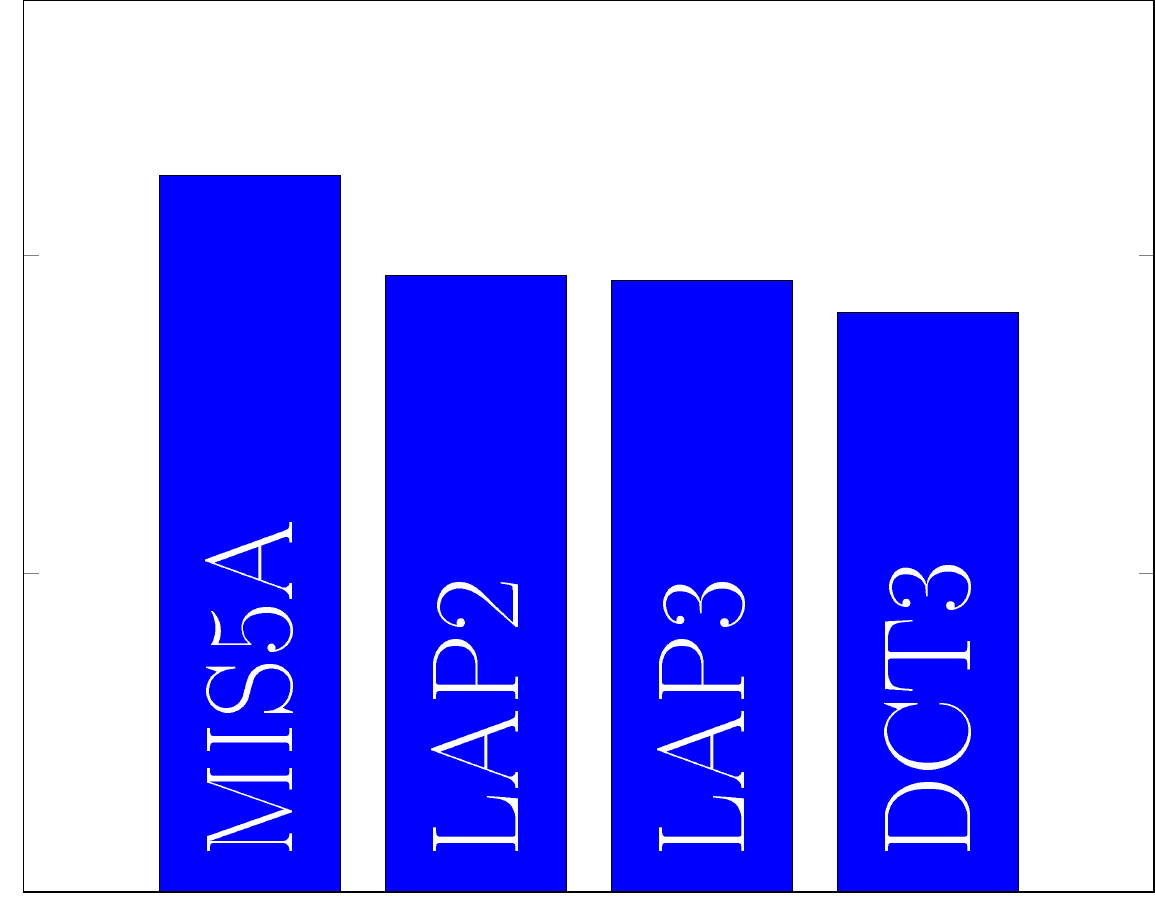} & 
\includegraphics[width=0.175\textwidth]{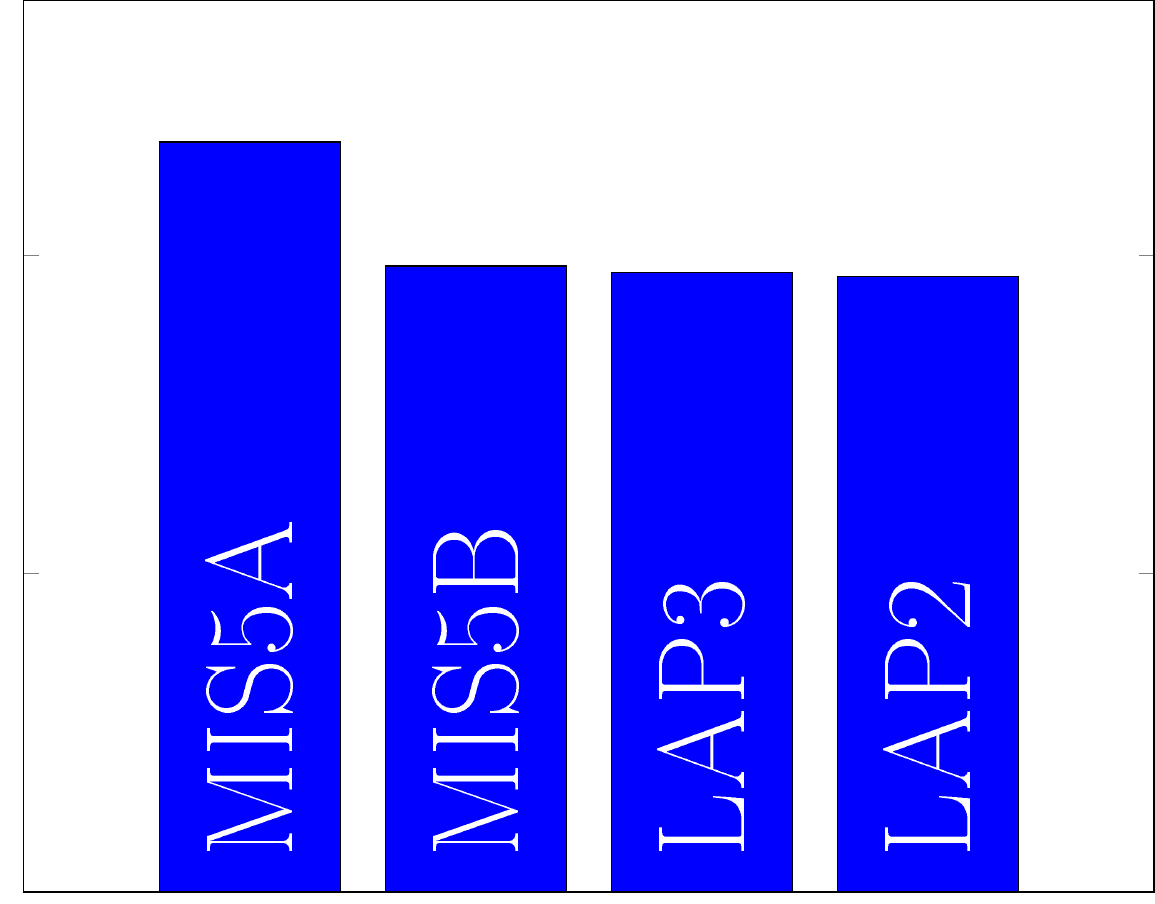} & 
\includegraphics[width=0.175\textwidth]{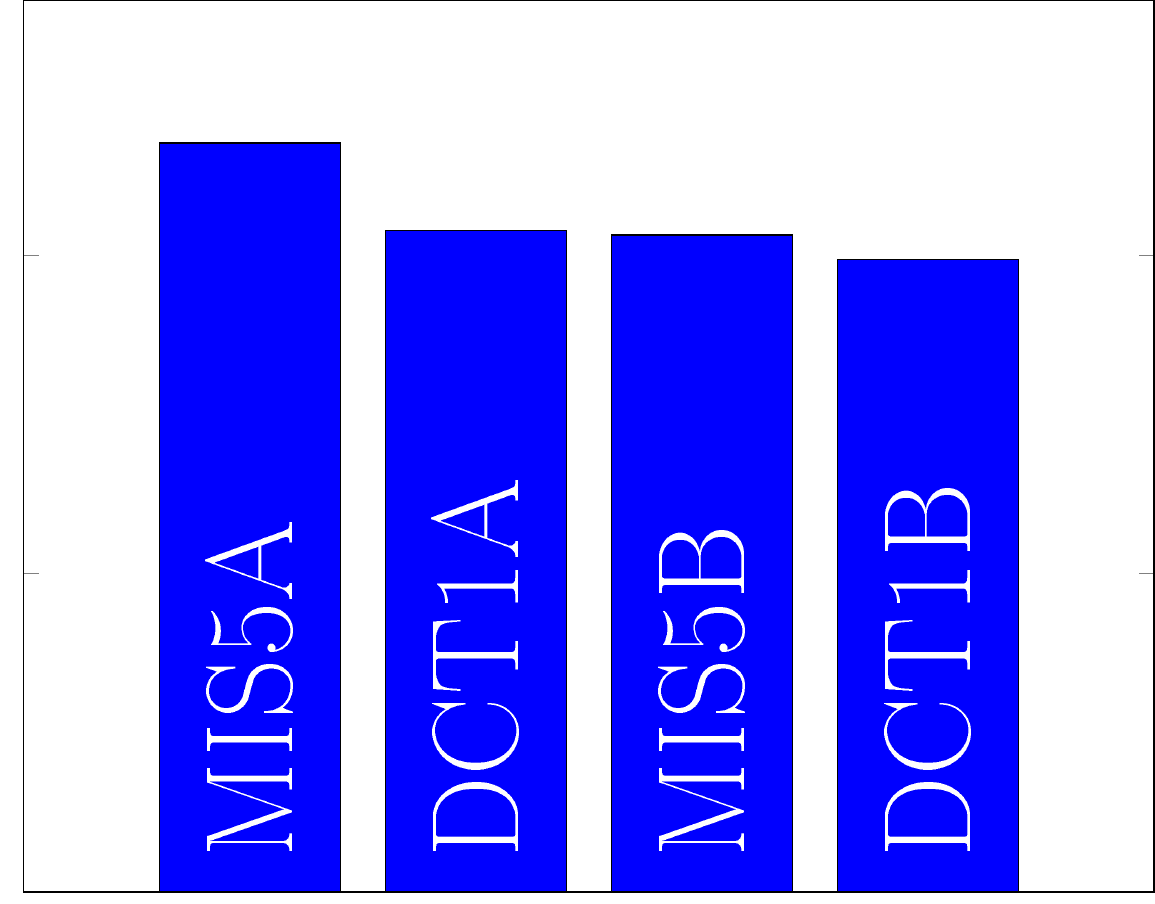} \\ \hline \hline
      &  & $D/r_0=6$ & $D/r_0=7$ & $D/r_0=8$ & $D/r_0=9$ & $D/r_0=10$ \\ \hline

 \raisebox{2.8\normalbaselineskip}[1.85cm][0pt]{\rotatebox[origin=c]{90}{ W1--W3}} &
  \raisebox{2.8\normalbaselineskip}[1.85cm][0pt]{\rotatebox[origin=c]{90}{ active regions}}  &\includegraphics[width=0.175\textwidth]{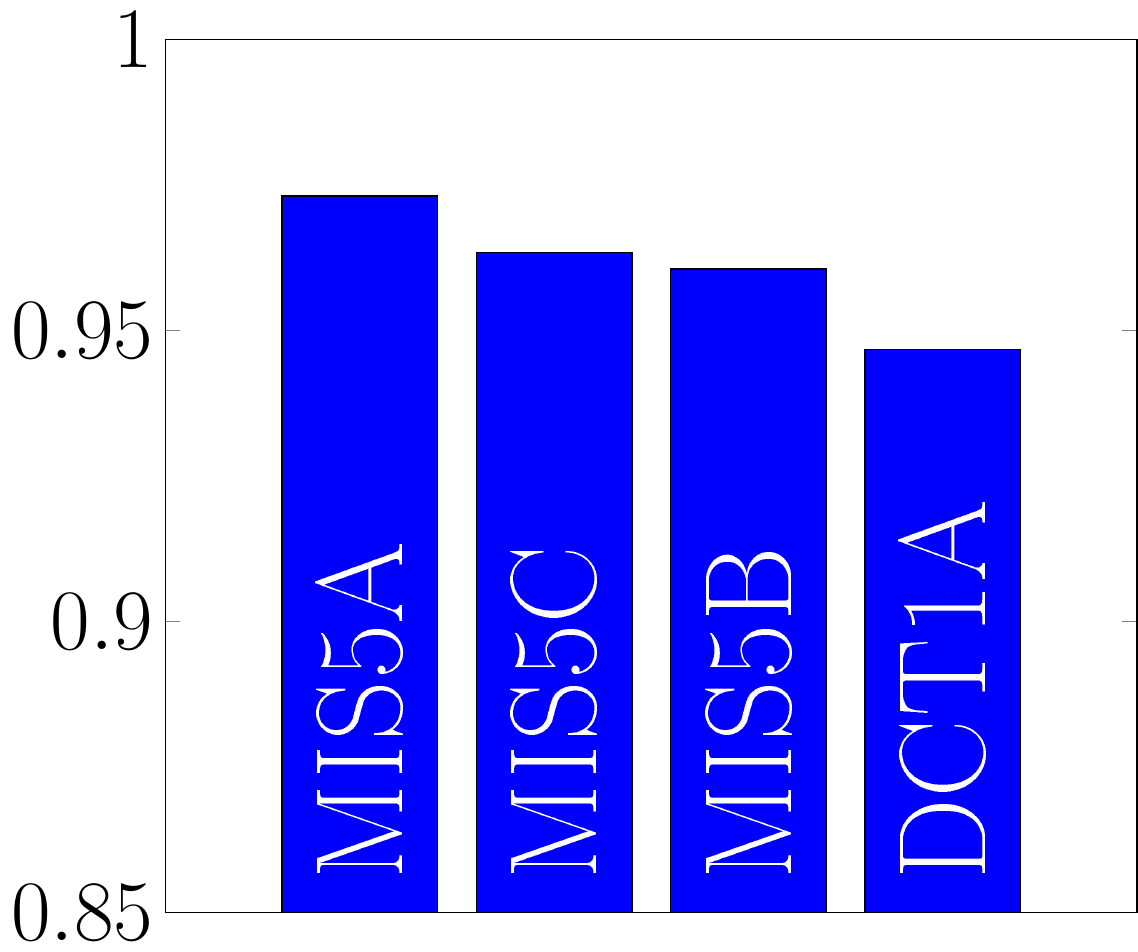} &  \includegraphics[width=0.175\textwidth]{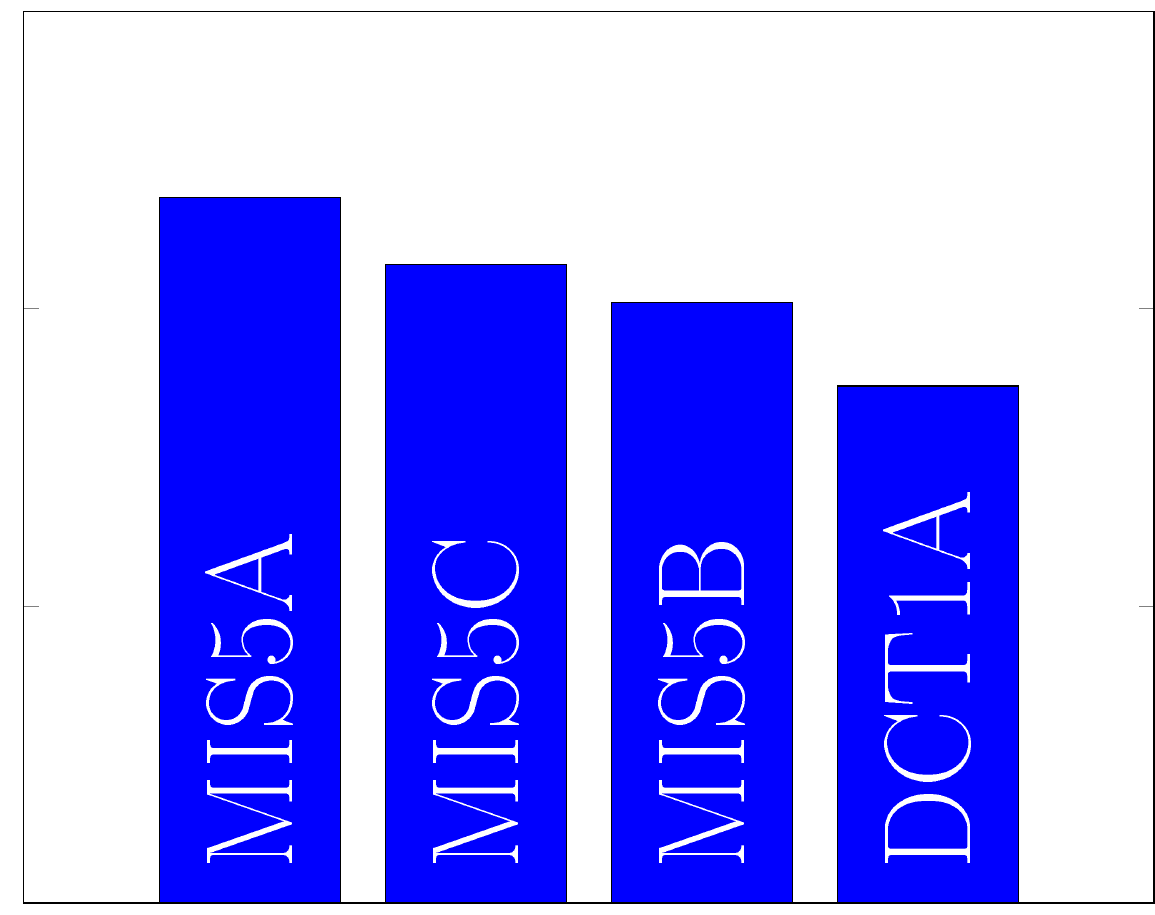} & 
\includegraphics[width=0.175\textwidth]{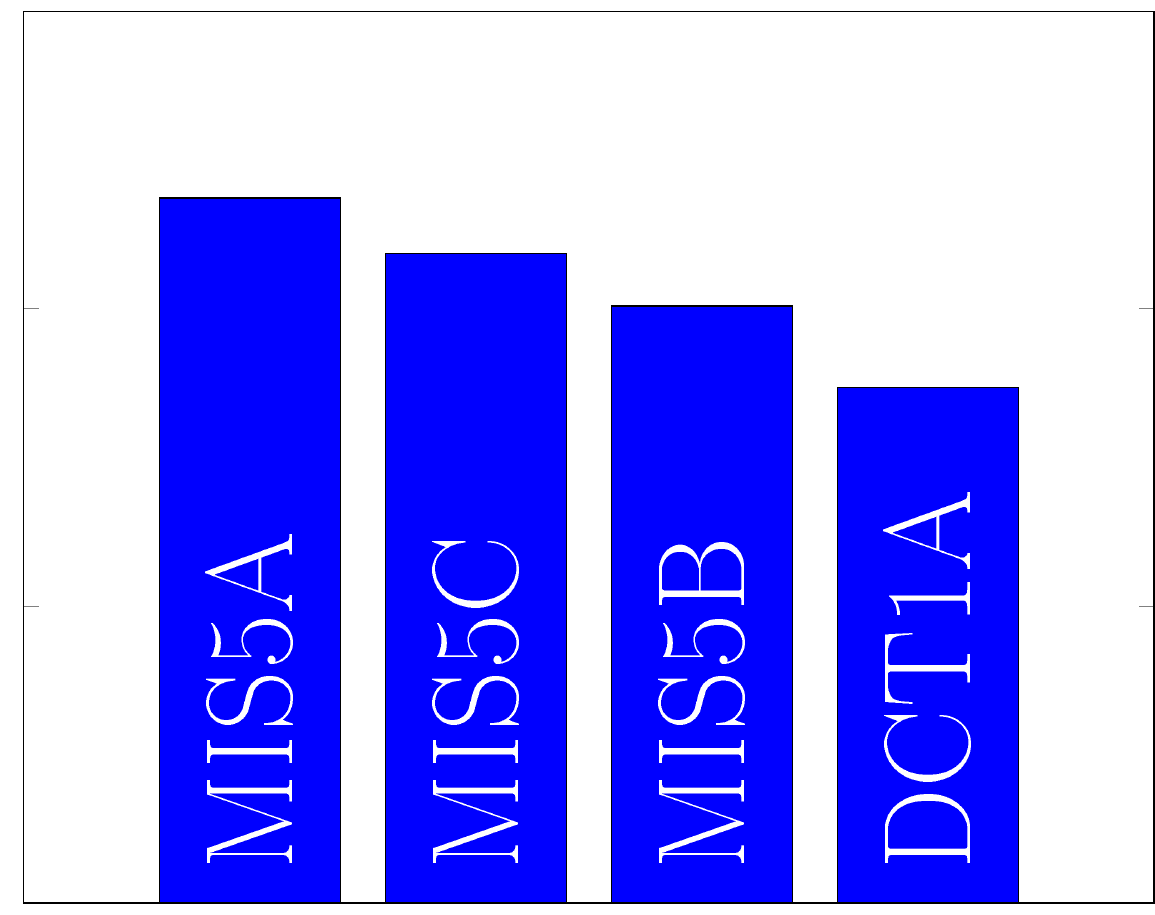} & 
\includegraphics[width=0.175\textwidth]{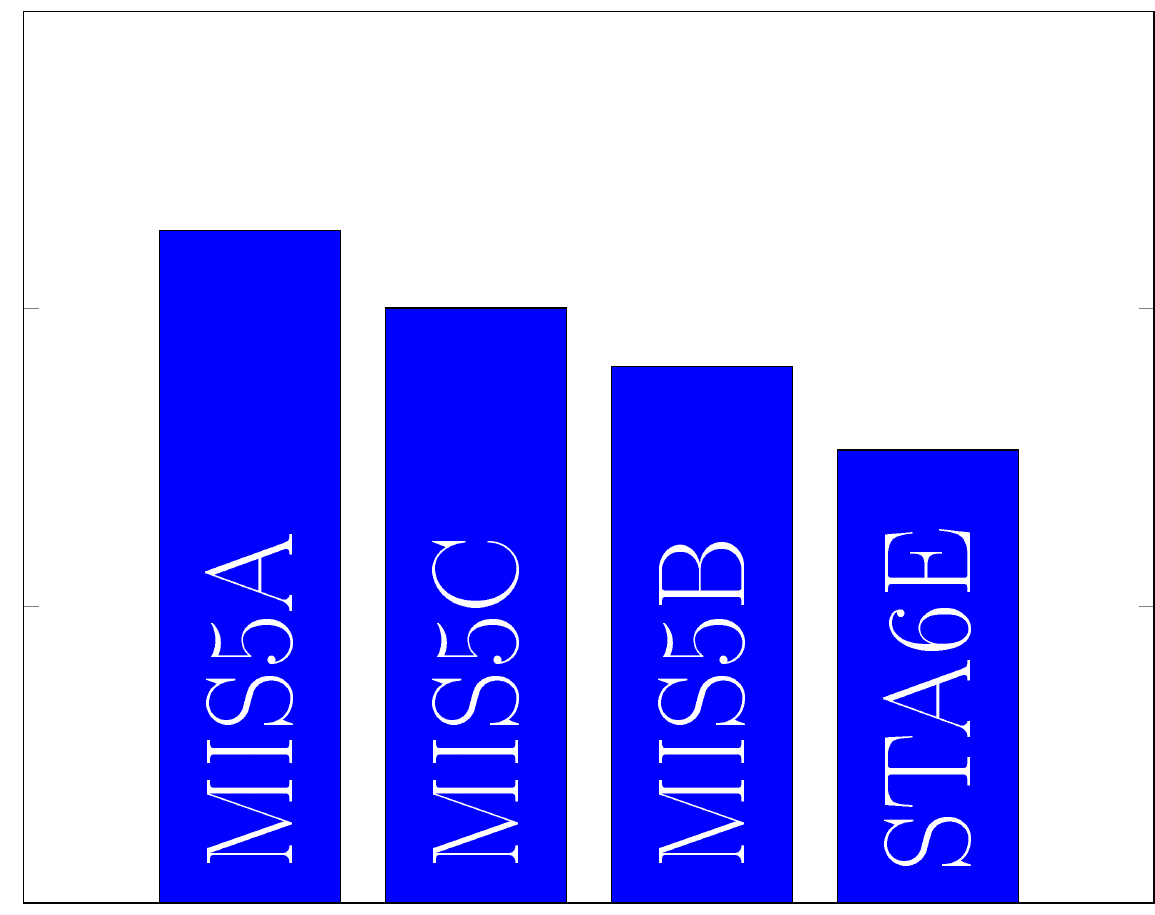} & 
\includegraphics[width=0.175\textwidth]{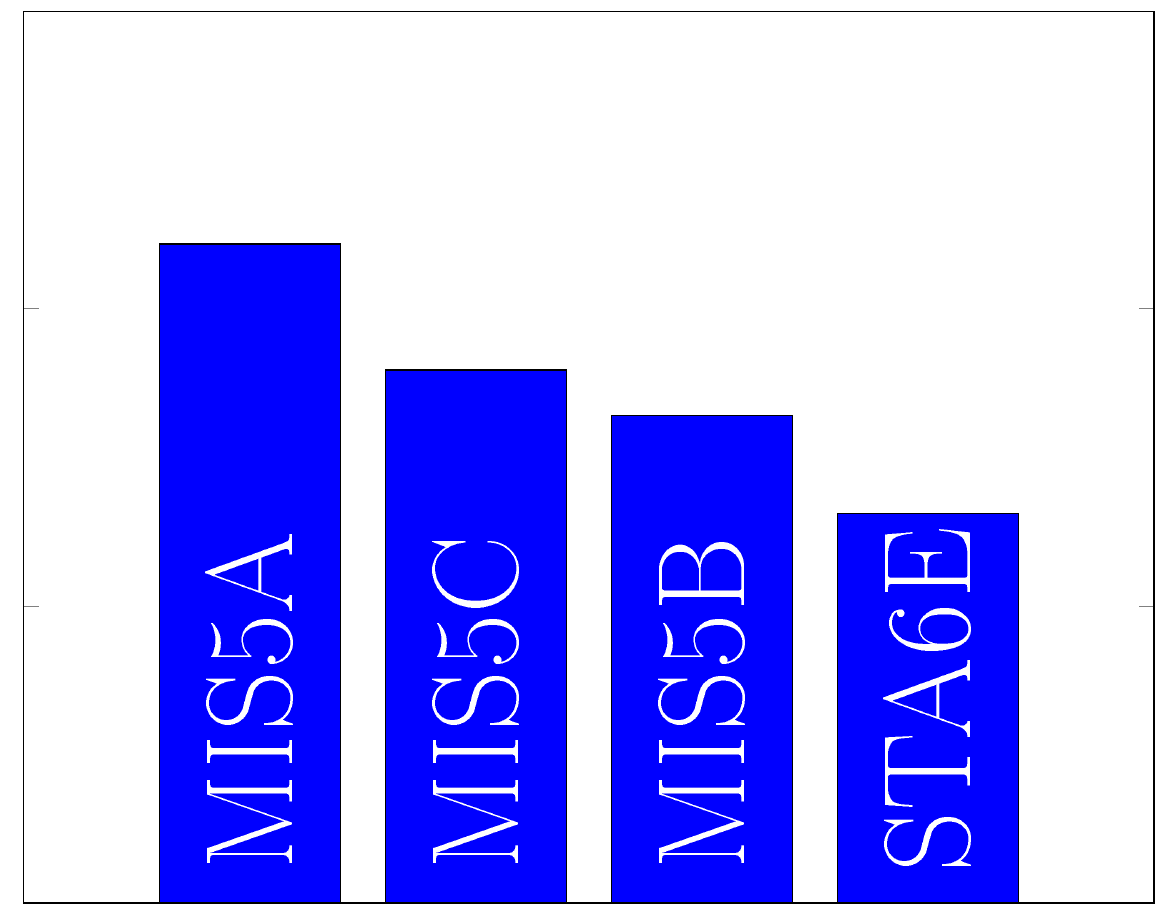} \\
 \raisebox{2.8\normalbaselineskip}[1.85cm][0pt]{\rotatebox[origin=c]{90}{ W4--W6}}  &
  \raisebox{2.8\normalbaselineskip}[1.85cm][0pt]{\rotatebox[origin=c]{90}{ granulation}}  &
\includegraphics[width=0.175\textwidth]{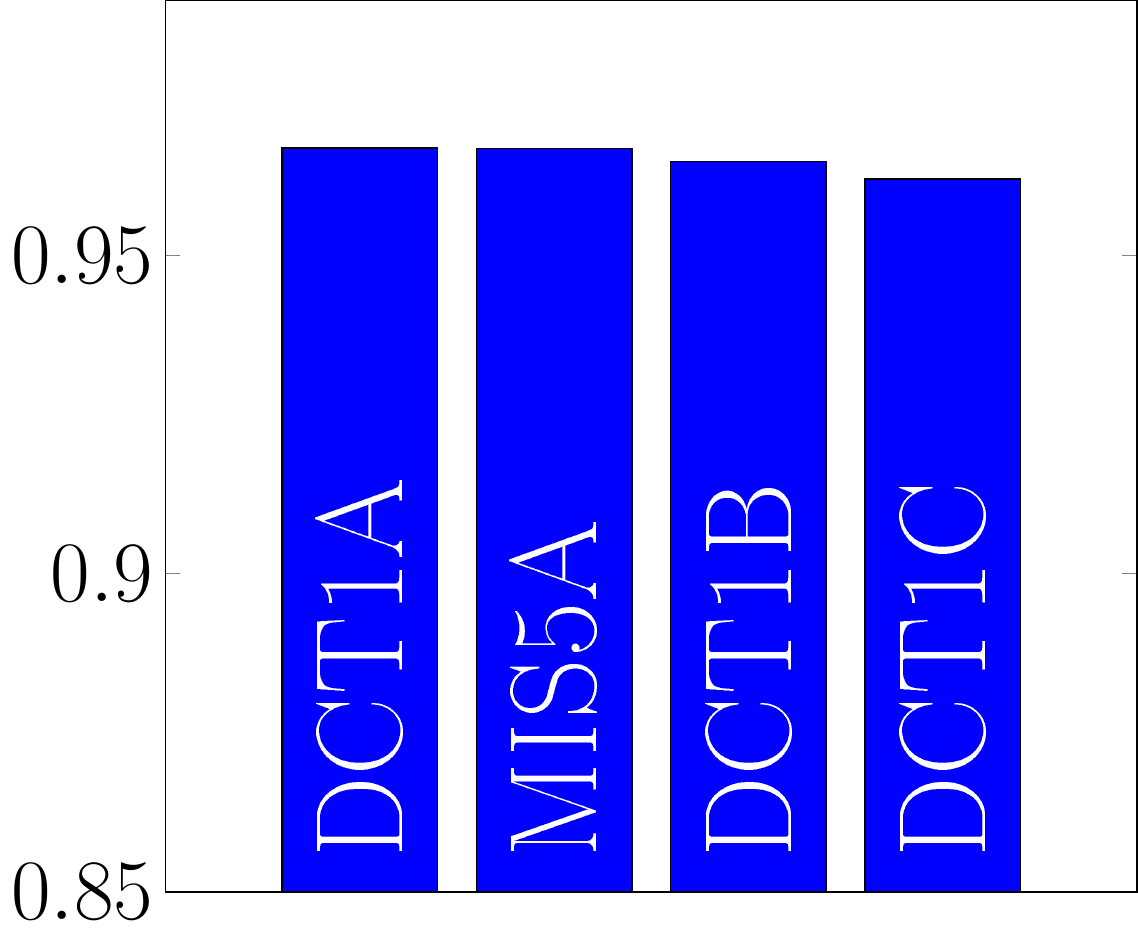} &  \includegraphics[width=0.175\textwidth]{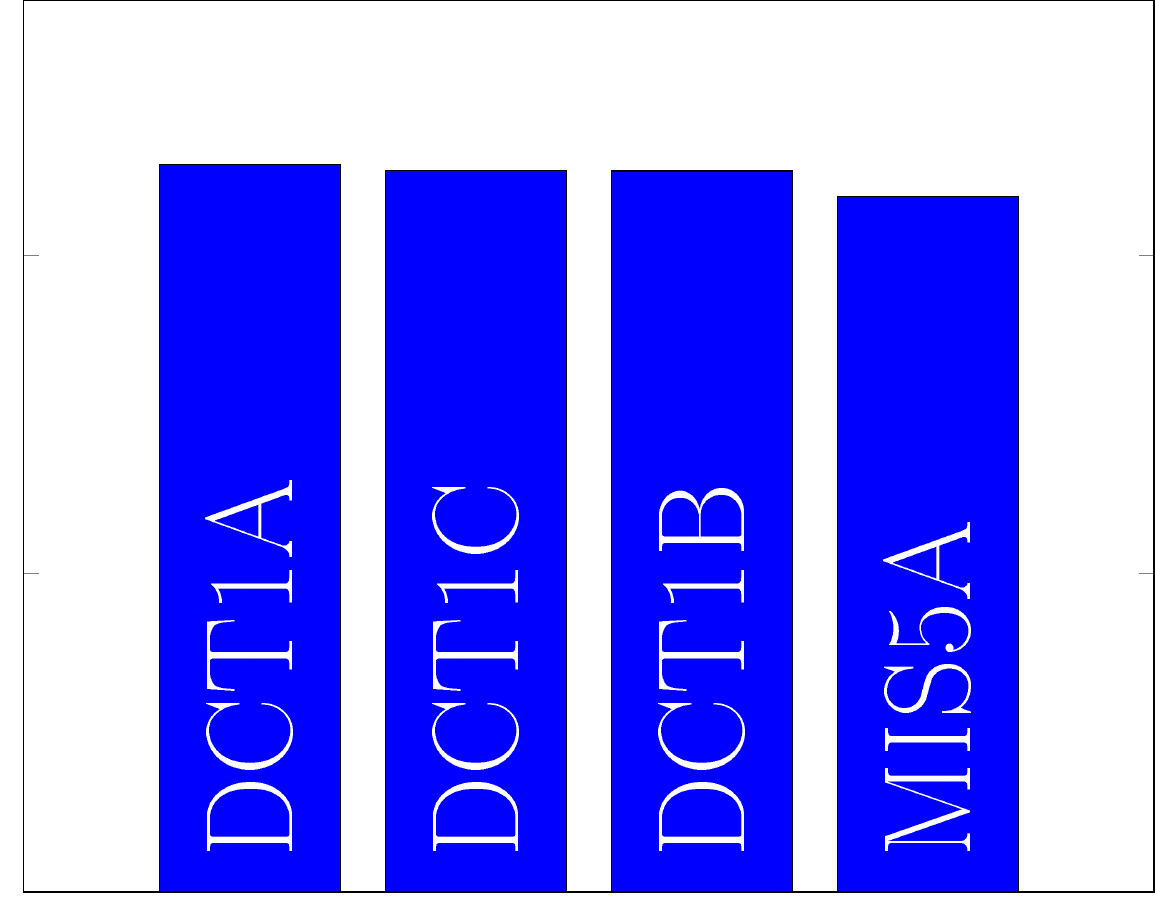} & 
\includegraphics[width=0.175\textwidth]{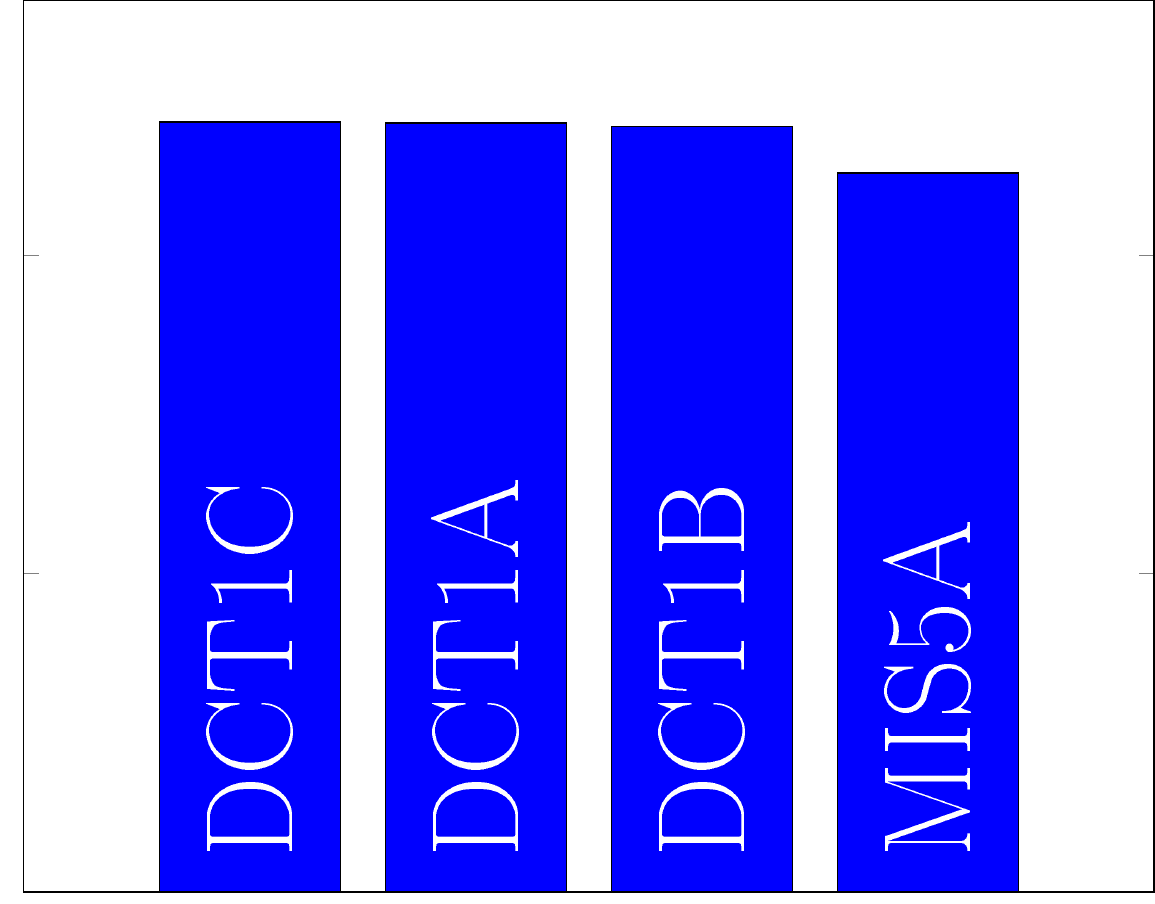} & 
\includegraphics[width=0.175\textwidth]{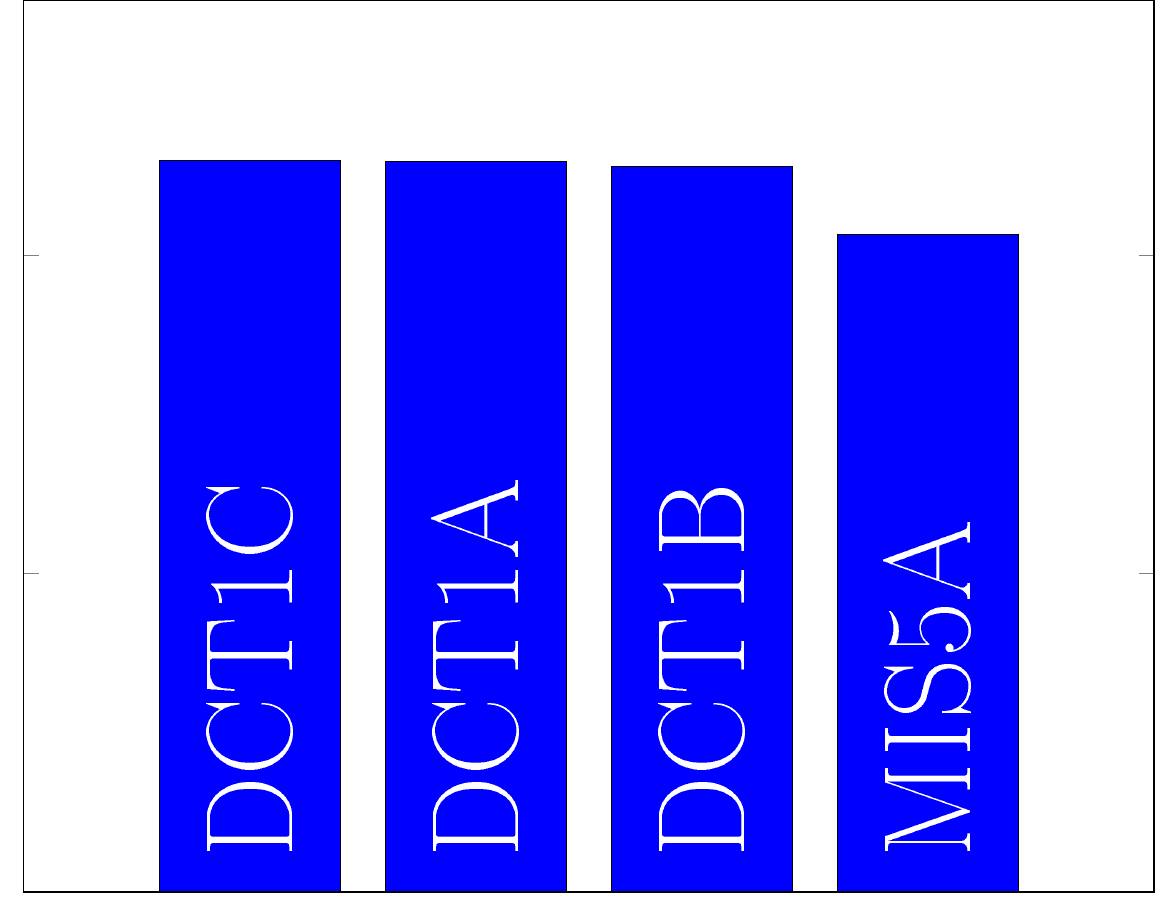} & 
\includegraphics[width=0.175\textwidth]{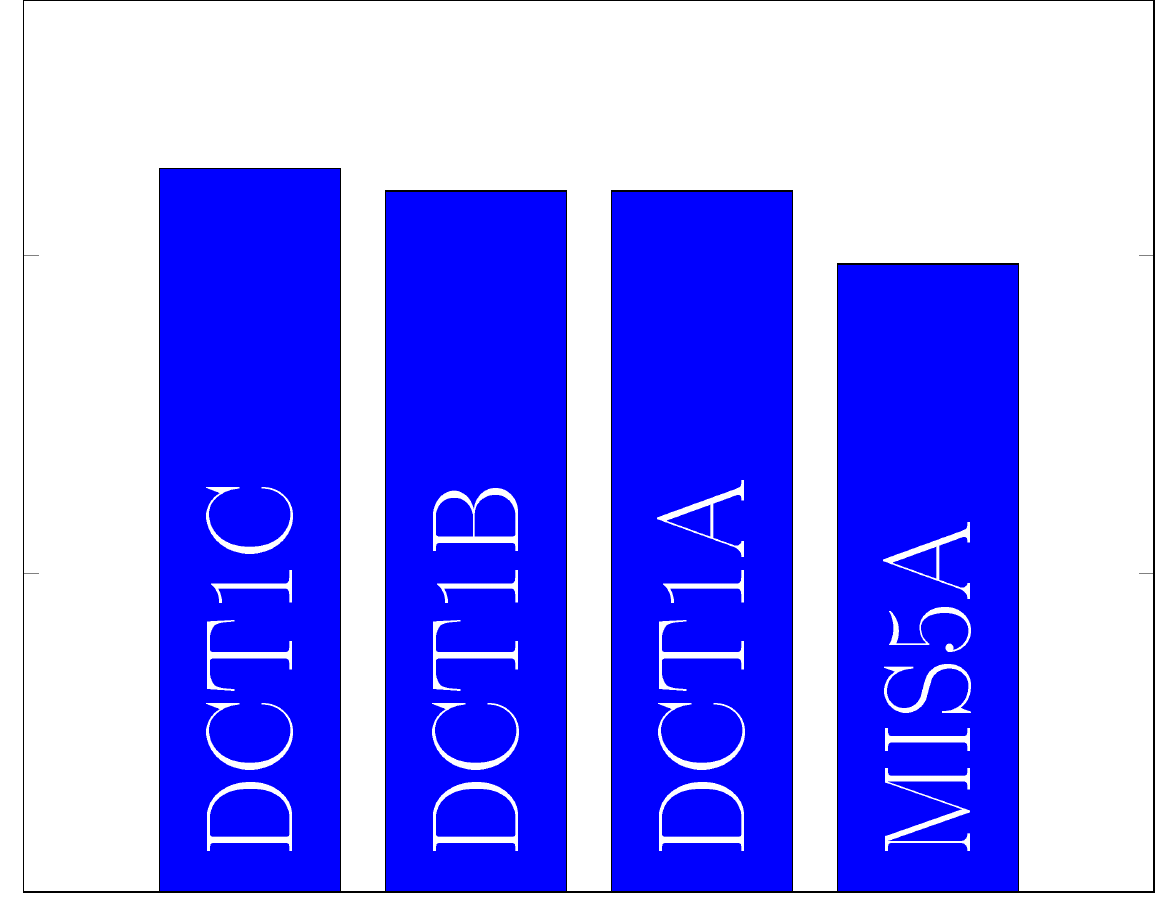} \\
\end{tabular}
\begin{tabular}{cc}
\hline \hline
\multicolumn{2}{c}{$D/r_0=1-10$} \\ \hline 
W1--W3 & W4--W6 \\
active regions & granulation \\
\includegraphics[height=5cm,width=0.475\textwidth]{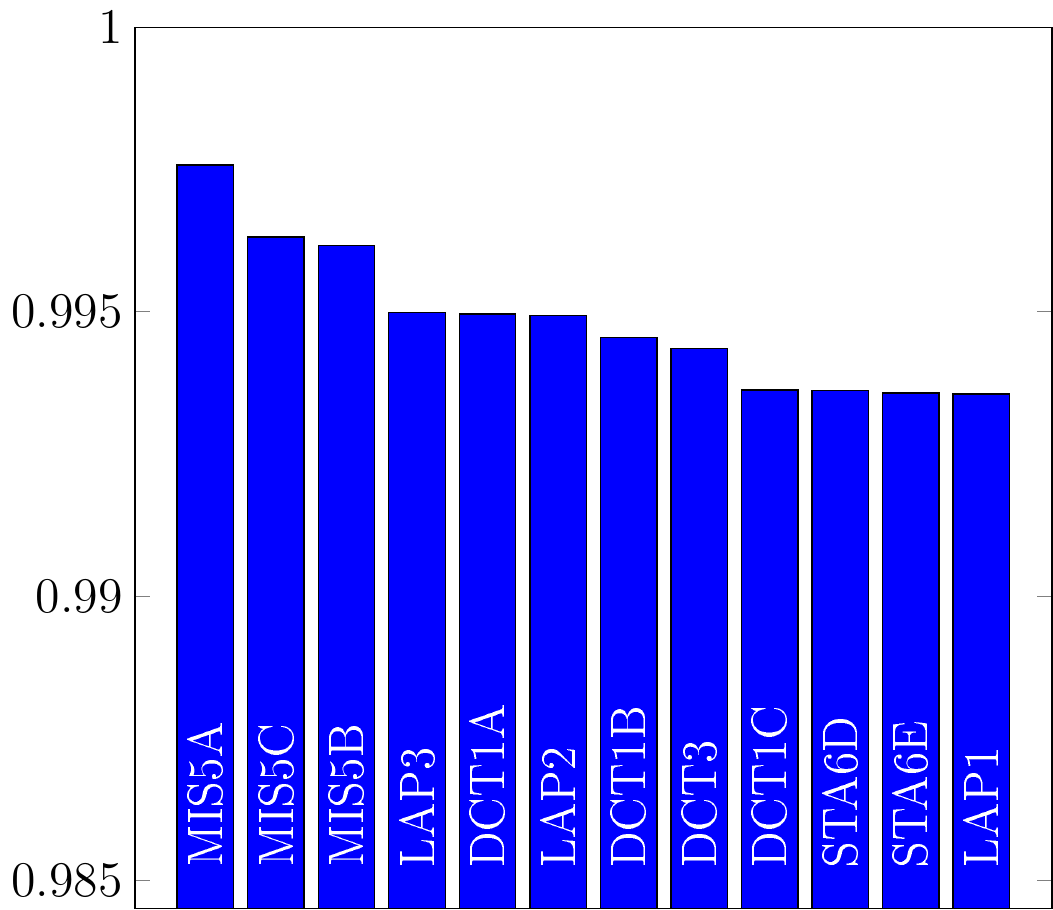} &
\includegraphics[height=5cm,width=0.475\textwidth]{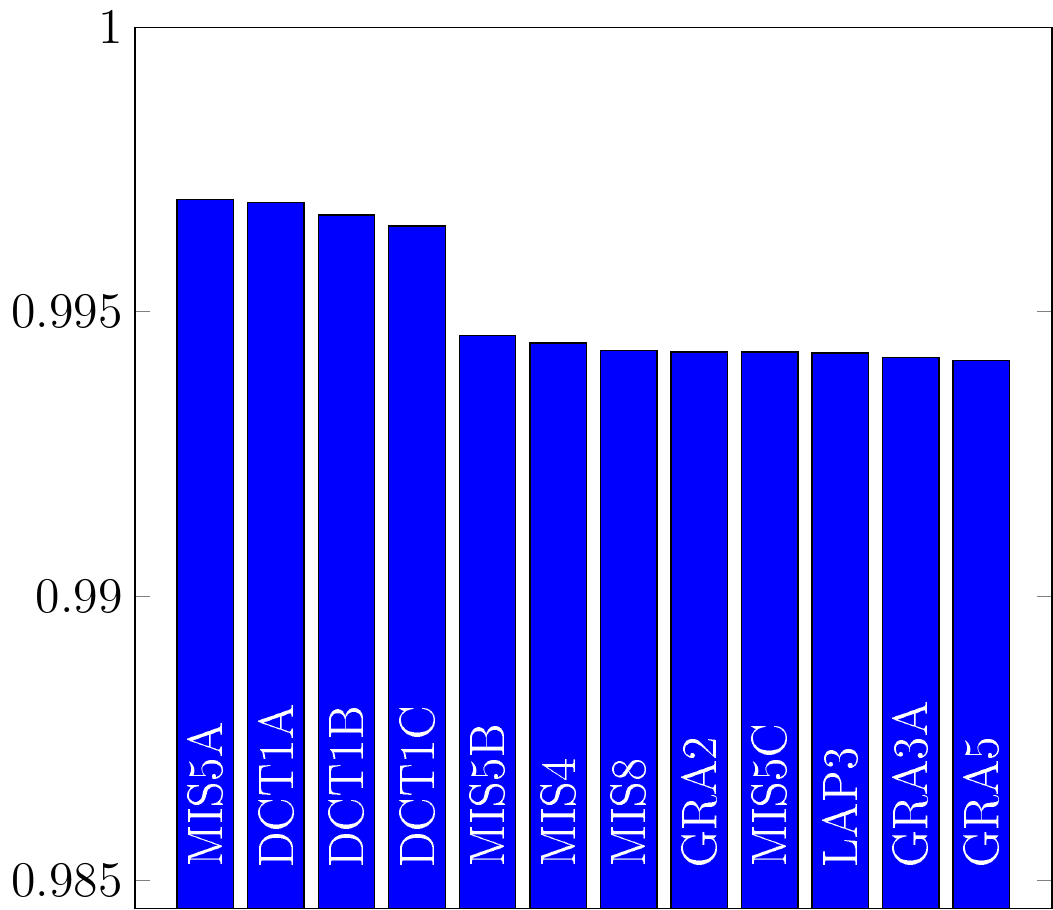}
 \\\hline 

\end{tabular}
\label{tab_wyniki}
\end{table*}

According to the results presented in Tab. \ref{tab_wyniki}, there is no single winner covering all possible atmospherics conditions and scene characteristics. However, there are three techniques which showed distinctively better performance and, therefore, they are discussed below. 

One of the methods with very impressive performance, is Helmli and Scherer's Mean method (MIS5) proposed by \citet{HelmliScherer2001}. This technique provided very good results for whole range of $D/r_0$ conditions and both types of observed solar regions. For all atmospheric conditions it is always among the four best methods. As indicated by distinctively higher correlation coefficients calculated over $D/r_0=1-10$, its usefulness should be considered when the seeing is highly varying or unknown. In summary, this method should be the first choice among all tested techniques.

The second noteworthy method is the Median Filter Gradient Similarity (GRA8) method, which was recently proposed by \citet{DengZHang2015}. It shows the best performance for very good atmospherics conditions ($D/r_0<4$). However, to achieve the high effectiveness we had to slightly modify the method by (1) combining both horizontal and vertical gradients and (2) changing the size of kernel. The superiority of the method is evident especially for active regions (W1$-$W3), while it is slightly less efficient when assessing only granulation patches (W4$-$W6), especially for $D/r_0>2$. Interestingly, this technique was not included in the best 12 methods indicated by wide-range correlations, $D/r_0=1-10$ which indicates that it should not be applied for observations with poor or unknown atmospherics conditions. The results of the GRA8 method recommend it for the utilization in observations assisted by AO since in this case the image quality is significantly enhanced, and the effective $D/r_0$ is small.

The last method, which provides good results is the DCT Energy Ratio proposed by \cite{ShenChen2006}. Its performance is the highest for moderate and poor atmospherics conditions when observing granulation regions. This method is second best method when the whole range of $D/r_0$ is taken into account, for patches W4-W6. The parameter of this method - the size of sub-block - should be selected accordingly to the blurring strength, i.e., the larger the $D/r_0$, the larger the sub-block.

\begin{figure}
\centering
\includegraphics[width=0.8\linewidth]{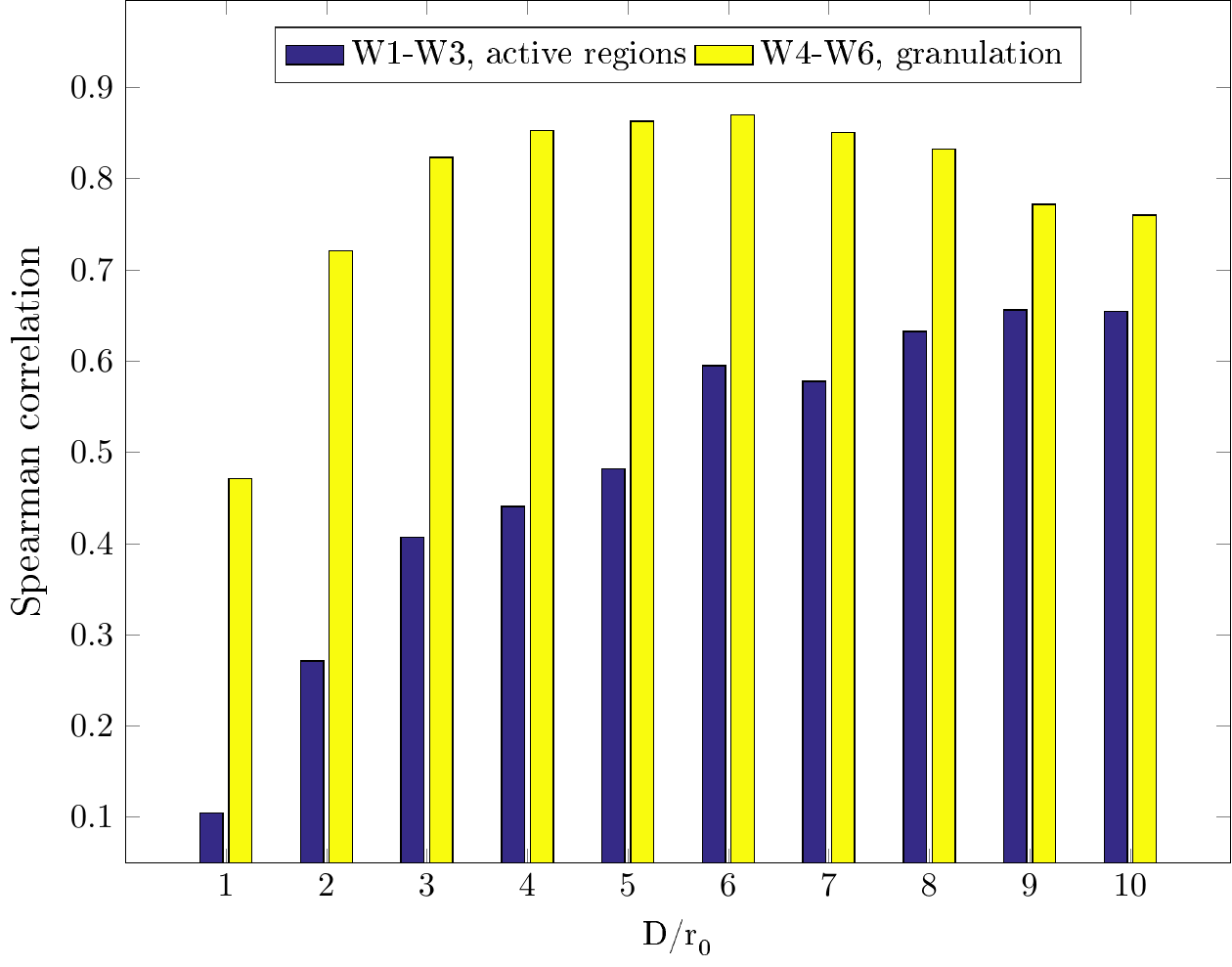}
\caption{Comparison of the effectiveness of most popular image quality technique used for solar images, i.e., rms-contrast. Significantly higher correlation values are achieved for granulation regions. }
\label{rms_test}
\end{figure}

The method frequently used in solar observations, rms-contrast or normalized gray-level variance (STA5), showed surprisingly poor performance. Since the rms-contrast measure was originally developed for granulation regions, with isotropic and uniform characteristics, we investigated how the method performed for this type of scene. The results presented in Fig. \ref{rms_test} indicate that indeed the techniques allows for achieving significantly higher correlation values for granulation. Still, the outcomes based on rms-contrast were much less correlated with the expected image quality than the results achieved by the best techniques. Therefore, it did not appeared in any ranking presented in Tab. \ref{tab_wyniki}. Our conclusions agree with the observations made by \cite{DengZHang2015} who also indicate relatively low efficiency of the common rms-contrast measure.

Some interesting conclusions can be drawn from the average performance of each QM family presented in Fig. \ref{families}. The Laplacian-based methods (LAP) show very good average correlation with $D/r_0$ for both types of observed scenes. While the LAP-based methods are indicated only several times in the rankings presented in Tab. \ref{tab_wyniki}, they still should be considered as a base for new, better techniques. The biggest difference between the efficiency when changing from granulation to active regions, can be observed in statistical methods (STA). This agrees well with the results shown in Fig. \ref{rms_test} and appears due to the assumption made in STA methods that regions should expose uniform properties across the observed field. The strange shape of GRA dependency visible in the right plot of Fig. \ref{families} arises due to the characteristics of  Median Filter Gradient Similarity (GRA8). As it was stated before, the performance of this method rises significantly as the $D/r_0$ decrease. This behavior biases the average efficiency of GRA family.

The possibility of using a method in real-time image selection can be especially important, as it allows for recording only the most useful data. Therefore, for completeness of our comparison, we measured the time required for processing a single image patch by each method. The evaluations were performed on a single core of IntelCore i7-3930K operating at 3.2~Ghz. The procedures were repeated $10^4$ times to obtain the average execution time. The results are presented in Tab.~\ref{tab_times} wherein we show the  methods and their average computation time (single-image analysis) in seconds, alternately. 

The results show that real-time frame selection for the assumed size of image patch is possible for frame rates more than 1000fps for most of the analyzed methods. This can be, however, not true if one wants to process large, full resolution, frames and/or if several steps of image calibration have to be applied. For such a case, the usage of graphics processing units (GPU) and accompanying optimization of a code should be considered. However, in our comparison we decided to perform measurements on a single CPU core, so that the further reduction of execution time with a multi-core machine can be estimated. For most of the included methods, the analysis is performed independently in local sub-regions of whole image, which makes them easily parallelized. 

We observed that the Median Filter Gradient Similarity (\emph{GRA8}) requires visibly more computation time ($>$0.01 s) than \emph{MIS5} ($<$0.002 s). It is mostly due to the median filtering which requires sorting pixels in a sliding window. This indicates that \emph{GRA8}, compared to \emph{MIS5}, would require more computing resources and/or better code optimization to operate at high frame rates utilized frequently in solar observations. Unfortunately, the Discrete Cosine Transform techniques (\emph{DCT1} and \emph{DCT2}) have significantly higher execution time, which makes them useless for real-time computation. These methods should be considered only in the post-processing. The fastest method was the one most frequently utilized in solar observations, \emph{STA5}. Unfortunately its mediocre performance is not compensated by the distinctively higher computational efficiency.

\begin{table*}
 \caption{Computation efficiency of the methods. In each consecutive column we show method abbreviation and execution time in seconds, alternately. The three  fastest methods are indicated in bold.}
 \begin{tabular}{llllllllllllll} 
 \hline  
MIS1A & 0.0020 & MIS7H & 0.0114 & GRA8C & 0.0388 & STA6C & 0.0026\\ 
MIS1B & 0.0025 & MIS8 & 0.0011 & GRA8D & 0.0752 & STA6D & 0.0021\\ 
MIS1C & 0.0029 & MIS9 & 0.0013 & GRA8E & 0.0082 & STA6E & 0.0023\\ 
MIS1D & 0.0039 & GRA1A & 0.0050 & GRA8F & 0.0081 & STA7A & 0.0013\\ 
MIS2A & 0.0010 & GRA1B & 0.0041 & LAP1 & 0.0017 & STA7B & 0.0018\\ 
MIS2B & 0.0011 & GRA1C & 0.0038 & LAP2 & 0.0027 & STA7C & 0.0031\\ 
\bf{MIS2C} & \bf{0.0006} & GRA1D & 0.0050 & LAP3 & 0.0039 & STA7D & 0.0046\\ 
MIS2D & 0.0023 & GRA1E & 0.0046 & LAP4 & 0.0014 & \bf{STA8A} & \bf{0.0006}\\ 
MIS2E & 0.0024 & GRA1F & 0.0044 & STA1A & 0.0012 & STA8B & 0.0021\\ 
MIS2F & 0.0026 & GRA2 & 0.0012 & STA1B & 0.0012 & STA8C & 0.0022\\ 
MIS2G & 0.0017 & GRA3A & 0.0009 & STA1C & 0.0012 & STA8D & 0.0020\\ 
MIS3 & 0.0160 & GRA3B & 0.0012 & STA1D & 0.0012 & DCT1A & 7.8562\\ 
MIS4 & 0.0065 & GRA3C & 0.0013 & STA1E & 0.0012 & DCT1B & 6.5742\\ 
MIS5A & 0.0028 & GRA3D & 0.0013 & STA2A & 0.0060 & DCT1C & 8.0760\\ 
MIS5B & 0.0031 & GRA3E & 0.0011 & STA2B & 0.0058 & DCT2A & 5.5642\\ 
MIS5C & 0.0027 & GRA3F & 0.0010 & STA2C & 0.0058 & DCT2B & 5.7970\\ 
MIS6A & 0.0768 & GRA3G & 0.0020 & STA2D & 0.0056 & DCT2C & 7.7149\\ 
MIS6B & 0.0855 & GRA3H & 0.0021 & STA2E & 0.0058 & DCT3 & 0.0011\\ 
MIS6C & 0.0806 & GRA3I & 0.0025 & STA3 & 0.0010 & WAV1A & 0.0099\\ 
MIS6D & 0.0809 & GRA3J & 0.0019 & STA4A & 0.0037 & WAV1B & 0.0143\\ 
MIS7A & 0.0135 & GRA3K & 0.0021 & STA4B & 0.0040 & WAV2A & 0.0105\\ 
MIS7B & 0.0136 & GRA4 & 0.0011 & STA4C & 0.0047 & WAV2B & 0.0150\\ 
MIS7C & 0.0141 & GRA5 & 0.0079 & STA4D & 0.0036 & WAV3 & 0.0265\\ 
MIS7D & 0.0129 & GRA6 & 0.0030 & STA4E & 0.0045 & WAVC & 0.0848\\ 
MIS7E & 0.0107 & GRA7 & 0.0023 & \bf{STA5} & \bf{0.0005} &  & \\ 
MIS7F & 0.0108 & GRA8A & 0.0072 & STA6A & 0.0019 &  & \\ 
MIS7G & 0.0106 & GRA8B & 0.0204 & STA6B & 0.0016 &  & \\ 
\hline
 \end{tabular} 

  \label{tab_times}
 \end{table*}

\section{Conclusions}

The assessment of image quality is an integral part of observations of the solar photosphere and chromosphere. It is essential to precisely select only the images of the highest fidelity before performing consecutive image reconstruction. Unfortunately the complexity and variety of observed structures make the estimation of image quality a challenging task. There is no single point-like object available, therefore, the selection techniques utilized in nighttime lucky imaging are not applicable.

In this study we decided to employ 36 techniques with various implementations and evaluated their precision in selecting the best exposures. The methods were based on various principles (gradients, intensity statistics, wavelets, Laplacians, Discrete Cosine Transforms) and were published over the last 40 years. Additionally, we enhanced their performance by applying simple modifications and by adjusting their tunable parameters.

In the comparison we employed reference images, containing both active regions and granulation areas, obtained by the Hidnoe satellite. The selected patches were degraded by convolution with blurring kernels generated by RWV method which faithfully reflects the seeing characteristics. We assumed a wide range of relative atmospheric conditions, starting from nearly undisturbed observations at $D/r_0=1$ to mediocre seeing at $D/r_0=10$. The reference quality of each simulated image was objectively estimated by measuring the amount of energy preserved in the Fourier spectrum of the original image. Eventually, to assess the efficiency of the compared techniques, we calculated Spearman's correlation coefficient between the outcomes of each method and the expected image quality.

The results of our comparison showed that the efficiency depends on the strength of atmospheric turbulence. For good seeing, $D/r_0<4$, the best method is the Median Filter Gradient Similarity \citep{DengZHang2015}, (GRA8), the most recent technique proposed for solar observations. Importantly, the original idea had to be slightly modified to enhance the method's performance. On the other hand, when the seeing conditions cover a wide range, the most efficient method is the Helmli and Scherer's Mean \citep{HelmliScherer2001}, (MIS5). This method should be considered when observing without AO or if the seeing conditions are unknown. The last distinctive method was the DCT Energy Ratio \citep{ShenChen2006} which showed high performance in poor atmospheric conditions when observing granulation regions. The measurements of execution time indicated that of the three mentioned techniques the Helmli and Scherer's Mean has significantly higher computational efficiency which recommends it for utilization with extremely high frame rates and/or larger image patches.

\begin{acks}
We would like to thank for numerous important comments, thorough review and great help in polishing the English provided by the anonymous Reviewer. This research was supported by the  Polish National Science Centre grants 2016/21/D/ST9/00656 and  (A. Popowicz) and 2015/17/N/ST7/03720 (K. Bernacki). We would like also to acknowledge the support from Polish Ministry of Science and Higher Education funding for statutory activities.
\end{acks}

\bibliographystyle{spr-mp-sola}
\bibliography{mybib}

\end{article} 

\end{document}